\documentclass[prb, aps, 10pt, twocolumn, floatfix, superscriptaddress]{revtex4-1}

\usepackage{amssymb,amsfonts,amsmath,amsthm}
\usepackage{graphicx}
\usepackage[english]{babel}
\usepackage[latin1]{inputenc}
\usepackage[pdftex,colorlinks=true,pdfstartview=FitH,linkcolor=blue,citecolor=blue,urlcolor=blue]{hyperref}

\usepackage{subcaption}

\usepackage{cancel}

\begin{document}

\title{
Cavity Optomechanics with a Laser Engineered Optical Trap
}

\author{P. Sesin}
\affiliation{Centro At\'omico Bariloche \& Instituto Balseiro (CNEA) and CONICET, Universidad Nacional de Cuyo (UNCuyo), Av. E. Bustillo 9500, R8402AGP S.C. de
Bariloche, RN, Argentina.}

\author{S. Anguiano}
\affiliation{Centro At\'omico Bariloche \& Instituto Balseiro (CNEA) and CONICET, Universidad Nacional de Cuyo (UNCuyo), Av. E. Bustillo 9500, R8402AGP S.C. de
Bariloche, RN, Argentina.}

\author{A.~E. Bruchhausen}
\affiliation{Centro At\'omico Bariloche \& Instituto Balseiro (CNEA) and CONICET, Universidad Nacional de Cuyo (UNCuyo), Av. E. Bustillo 9500, R8402AGP S.C. de
Bariloche, RN, Argentina.}

\author{A. Lema\^itre}
\affiliation{Centre de Nanosciences et de Nanotechnologies, C.N.R.S., Universit\'e Paris-Sud, Universit\'e Paris-Saclay, 10 Boulevard Thomas Gobert, 91120 Palaiseau, France.}

\author{A. Fainstein}
\email[Corresponding author, e-mail: ]{afains@cab.cnea.gov.ar}
\affiliation{Centro At\'omico Bariloche \& Instituto Balseiro (CNEA) and CONICET, Universidad Nacional de Cuyo (UNCuyo), Av. E. Bustillo 9500, R8402AGP S.C. de
Bariloche, RN, Argentina.}

\date{\small \today}

\begin{abstract}
Laser engineered exciton-polariton networks could lead to dynamically configurable integrated optical circuitry and quantum devices. Combining cavity optomechanics with electrodynamics in laser configurable hybrid designs constitutes a platform for the vibrational control, conversion, and transport of signals. With this aim we investigate 3D optical traps laser-induced in quantum-well embedded semiconductor planar microcavities.  We show that the laser generated and controlled discrete states of the traps dramatically modify the interaction between photons and phonons confined in the resonators, accessing through coupling of photoelastic origin $g_\mathrm{0}/2\pi\sim$1.7\,MHz an optomechanical cooperativity $C>1$ for mW excitation. The quenching of Stokes processes and double-resonant enhancement of anti-Stokes ones involving pairs of discrete optical states in the side-band resolved regime, allows the optomechanical cooling of 180\,GHz bulk acoustic waves, starting from room temperature down to $\sim$120\,K. These results pave the way for dynamical tailoring of optomechanical actuation in the extremely-high-frequency range (30-300\,GHz) for future network and quantum technologies.
\end{abstract}


\maketitle

\section{Introduction} 

Trapping and potential landscape engineering through non-resonant laser excitation is used  in the exciton-polariton domain to modify the polaritons spatial distribution, their spectra, and dynamics, including the formation of one and two-dimensional lattices, with prospects for new devices and quantum simulators \cite{Schneider,Wertz,Tosi,Abbarchi,Dreismann,Pieczarka,Anguiano,Boozarjmehr,Alyatkin}. In these experiments the applied laser defines an effective potential for the exciton polaritons, actuating either on their excitonic or photonic component. The physical mechanism that acts on the excitonic component is the Coulomb repulsion, which increases the energy of the coupled particles \cite{Wertz,Tosi,Abbarchi,Dreismann,Pieczarka}. The action on the photonic component depends on local laser-induced variations of the refractive index, usually defining attractive potentials \cite{Anguiano,Boozarjmehr}. This latter strategy can be also used to generate confining cavities, or more arbitrary effective potentials, for pure photons \cite{Anguiano}.

Lasers are routinely used in atomic physics to engineer optical traps for atoms and, in addition, to laser-cool them by quenching the atom motion through so-called Doppler cooling \cite{Letokhov,Wineland,Monroe,Chang}. Laser trapping by optical tweezers with 
vibrational cooling or stimulation have also been demonstrated recently for centre-of-mass oscillations of nanoparticles 
levitated in vacuum.~\cite{Pettit, Windey, Delic} Concepts related to Doppler cooling are also applied in a variety of solid systems for optomechanical dynamical backaction phenomena \cite{ReviewCOM}, including cooling of mechanical vibrations (even down to the quantum limit) \cite{O'Connell,Teufel,Chan,Verhagen,Kippenberg}, and laser-induced mechanical self-oscillation \cite{Kippenberg2,Grudinin}.  To the best of our knowledge, however, such cavity optomechanical phenomena have not been reported in condensed matter laser engineered optical resonators.

In this work we demonstrate optically generated photonic traps as optomechanical devices. A planar semiconductor microcavity with a thick spacer constituted by a superlattice (SL) with 41 GaAs/AlAs bilayers, is shown to one dimensionally confine near-infrared photons and vibrations with frequencies as high as $\sim$180\,GHz. 
Application of a green focussed laser beam induces in-plane trapping of photons with a full discretisation of the optical spectra, strongly enhancing the light-vibrational coupling.  We demonstrate single and double resonant Brillouin processes, and optomechanical cooling of the $\sim$180\,GHz slow phonons of the semiconductor SL from room temperature down to $\sim$120\,K.  The  laser-induced trap could be dynamically modified so that vibrations are selectively laser-cooled, or made to self-oscillate, opening the path for a variety of quantum information reconfigurable applications.

\section{Results}

\begin{figure*}[!hht]
 \begin{center}
\includegraphics*[keepaspectratio=true, clip=true, trim = 0mm 20mm 0mm 0mm, angle=0, width=1.9\columnwidth]{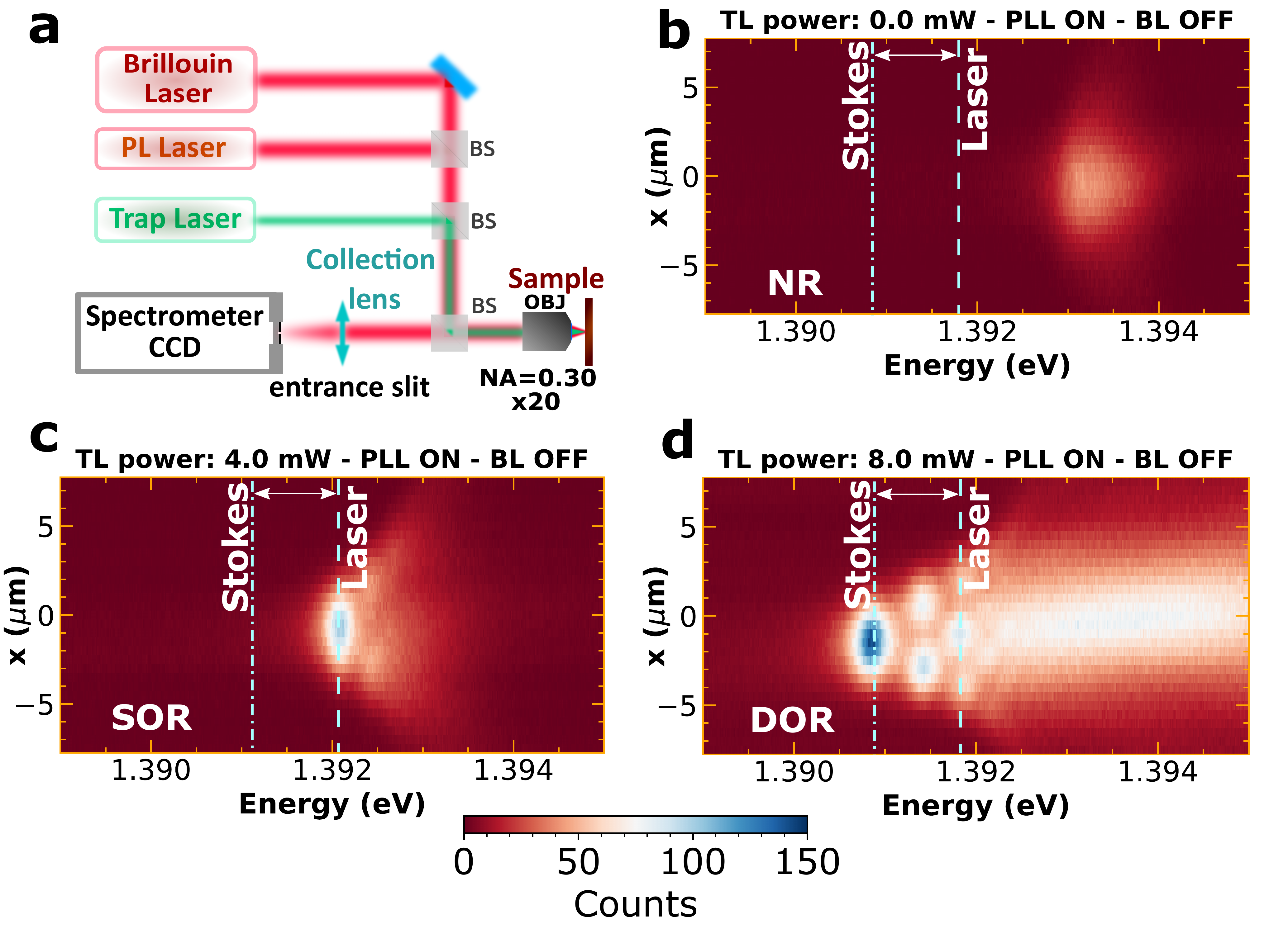}
\end{center}
\caption{
\textbf{Cavity optomechanics with laser-generated optical traps.}
\textbf{a} Scheme of the three-lasers experimental set-up used. \textbf{b-d} Spatial images of the  optical cavity modes that evolve with increasing trap laser (TL) power from planar 1D confinement in \textbf{b} to fully 3D confined in \textbf{c} and \textbf{d}. The energy of the Brillouin laser and Stokes scattered photons are indicated with vertical lines. In \textbf{b} the optomechanical process is non-resonant (NR). The situations corresponding to single (SOR) and double (DOR) optical resonances are attained in \textbf{c} and \textbf{d}, respectively. 
}
\label{Fig01}
\end{figure*}

\paragraph*{\textbf{Optomechanics with a laser induced optical trap.}} 

The concept for the proposed cavity optomechanical experiments with laser-trapped light (the set-up is schematized in Fig.\,\ref{Fig01}a and described in the Methods section below) is illustrated in Figs.~\ref{Fig01}b-d. These figures show spectrally resolved spatial images of the photoluminescence (PL) corresponding to increasing ``trap laser'' (TL) power (0, 4, and 8~mW, respectively). These images were obtained with the ``photoluminescence laser''  (PLL) on and the ``Brillouin laser'' (BL) off. The energy of the BL is indicated in the three panels with vertical dashed lines, together with that of the Stokes (red-shifted)  photons corresponding to the scattering by the $\sim$180\,GHz ($\sim$5.7\,cm$^{-1}$) slow bulk acoustic mode of the GaAs/AlAs superlattice embedded in the resonator.\cite{TrigoSurfaceAvoding,TrigoPRL,Villafane} This mode has almost zero-group velocity because it is located at the lower-edge of the Brillouin zone-center first phononic bandgap of the SL. Its slow speed leads to an effective mechanical quality factor $Q_m \sim 10^3$.  In Fig.~\ref{Fig01}b the broad emission corresponding to the planar microcavity can be observed starting at $\sim$1.3932\,eV and expanding towards higher energies (due to the parabolic dispersion of the 1D confined optical cavity modes). This situation corresponds to non-resonant (NR) inelastic scattering, in the sense that neither the BL nor the Stokes photons are resonant with any optical cavity mode. As the TL power is increased to 4~mW (Fig.~\ref{Fig01}c), clear modes of a 3D optical trap emerge, with the fundamental mode having redshifted with respect to the continuum, and falling precisely at the energy of the BL. This situation corresponds to a ``single optical resonance'' (SOR): the BL is resonant, but the Stokes photons are not. Finally, with the TL power at 8~mW (Fig.~\ref{Fig01}d), multiple discrete 3D optical trap modes are evidenced. The fundamental mode has further redshifted to be resonant with the Stokes photons, while the BL is resonant with the third optically confined mode. This situation corresponds to a ``double optical resonance'' (DOR). We note that for this experiment the separation between the discrete optically confined modes was tuned by choosing the size of the TL illuminated spot, which defines the lateral width of the Gaussian optical confining potential.

\begin{figure*}[t]
 \begin{center}
\includegraphics*[keepaspectratio=true, clip=true, trim = 0mm 0mm 0mm 0mm, angle=0, width=2.1\columnwidth]{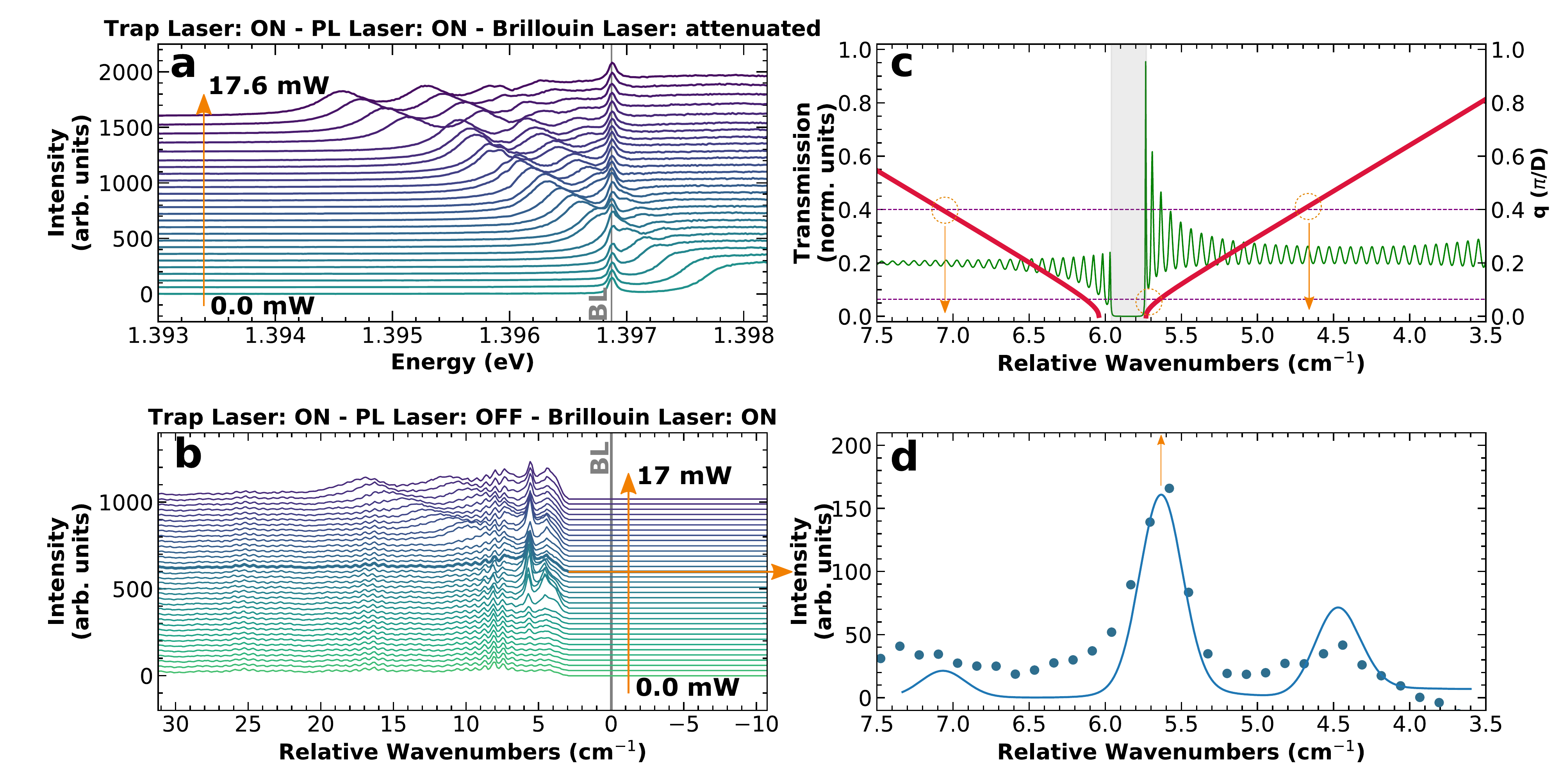}
\end{center}
\caption{
\textbf{Trap laser dependence of optical trap modes and optomechanical processes.}
 Photoluminescence (\textbf{a}) and Stokes Brillouin spectra (\textbf{b}) as a function of trap laser power.
In \textbf{a} the peak highlighted with a vertical line is the attenuated Brillouin laser (BL). In \textbf{b} the main peak at $\sim$5.7\,cm$^{-1}$ corresponds to the Stokes scattering by the slow $\sim$180\,GHz vibrational mode of the SL filling the cavity spacer (see text for further details).  \textbf{c} The calculated SL phonon dispersion is shown with the thick curve. It allows the identification of the $\sim$5.7\,cm$^{-1}$ peak with the slow Brillouin zone-center vibrational mode. The thin curve is the calculated phonon transmission for vibrations incident from the substrate, evidencing the presence of the SL Brillouin zone-center gap. 
\textbf{d} Comparison between the measured and calculated  SL Brillouin spectra. The width of the former is determined by the experimental resolution. This has been taken into account in the calculated spectrum with a Gaussian convolution.}
\label{Fig02}
\end{figure*}

Figure~\ref{Fig02}a displays the evolution of the 3D optical trap modes as they red-shift and emerge from the continuum when the depth of the optical trap is augmented with increasing TL power (from 0 to 17\,mW). For these spectra, taken in a sample position slightly different from the spatial images in Fig.\,\ref{Fig01},   the three lasers were ``on'', although the BL was strongly attenuated just to be able to detect its energy ($\sim$1.3968\,eV) superimposed on the emission spectra. While evolving with increasing TL power the 3D Gaussian trap optical modes successively cross the energy of the BL to be used in the following experiments (indicated by the vertical line in the figure). The 3D trap confined modes and the BL  are well below the SL gap ($\sim$1.44\,eV) in these experiments. Figure~\ref{Fig02}b presents the Brillouin spectra for the same sequence of TL powers. The spectra are given in relative wavenumbers, with the energy referred to that of the BL. In these experiments the PLL was blocked (``off''), and the power of the BL was increased to its working condition.  No signal is observed below 4\,cm$^{-1}$ due to the blocking of the laser light introduced by the first two-stages of the triple spectrometer (see the Methods section for experimental details). The stronger contribution to the spectra corresponds to the acoustic slow-mode of the SL at $\sim$5.7\,cm$^{-1}$, plus some additional minor contributions due to (i) propagating SL vibrational modes  almost symmetrically separated $\sim$1\,cm$^{-1}$ from the slow-mode (these phonons are observed due to the back-scattering of photons in the cavity, see the Supplementary Note 3 for further details), (ii) some narrow background oscillations that extend through the whole spectra, but are peaked periodically at $\sim$8.3, 16.6, 25\,cm$^{-1}$, and (iii) some weak photoluminescence presumably  excited by residual penetration of the TL. 
Figure~\ref{Fig02}d compares one of the experiments with a calculation of the Brillouin spectra, convoluted with the experimental resolution (see details of the calculation in the Supplementary Note 3). This, together with the calculated SL phonon dispersion shown in Fig.\,\ref{Fig02}c, clearly identifies the main peak in the spectra with the SL zone-center slow mode. Note that the intensity of this peak strongly varies as the optical trapping conditions are modified in Fig.\,\ref{Fig02}b. This is the first central result of this work, and will be discussed in detail next.

\begin{figure}[t!!!!]
 \begin{center}
\includegraphics*[keepaspectratio=true, clip=true, trim = 0mm 0mm 0mm 0mm, angle=0, width=\columnwidth]{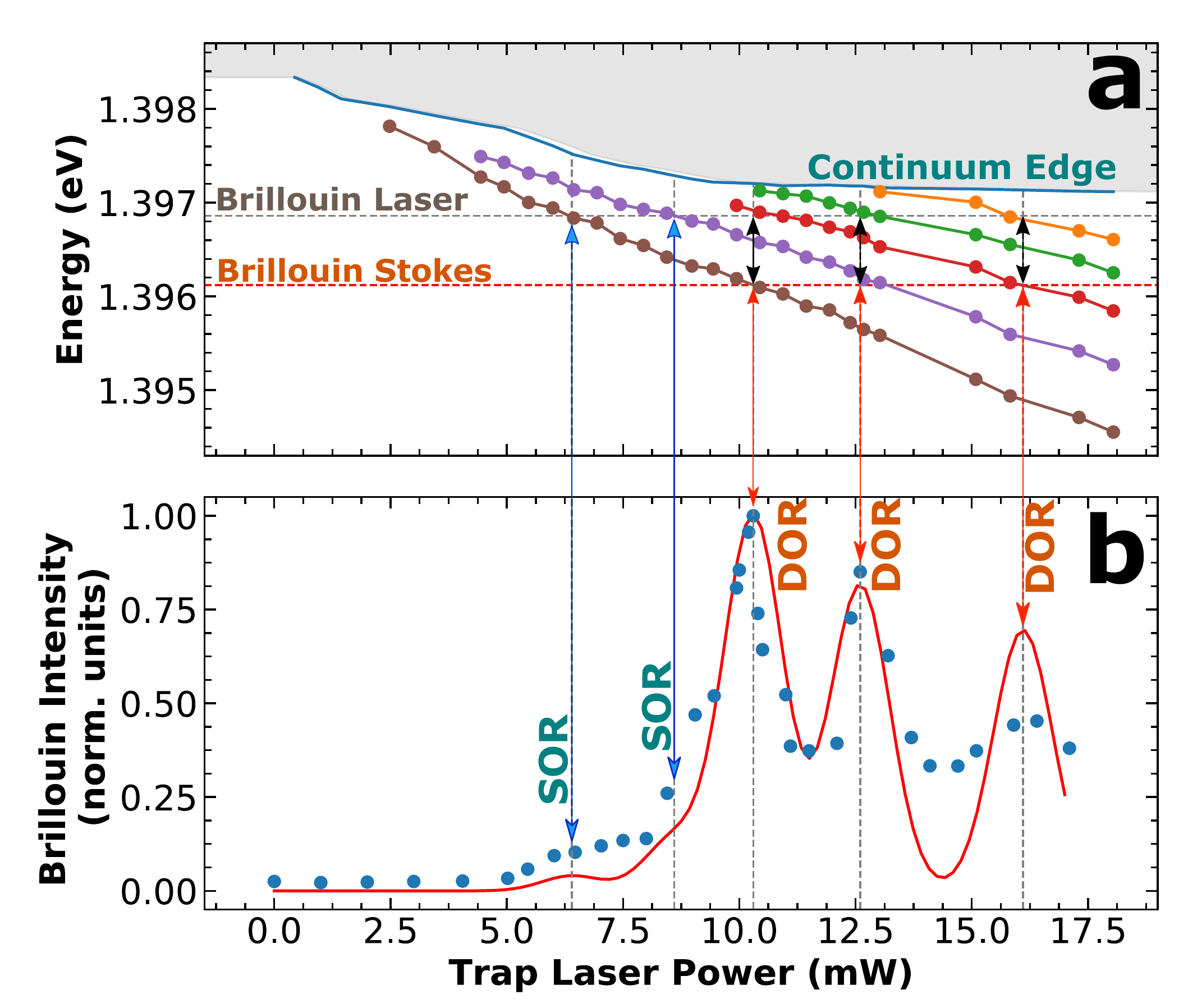}
\end{center}
\caption{
\textbf{Optomechanical coupling as a function of trap laser power.}
\textbf{a} Fan-plot of the energy of the optical trap modes as a function of TL power, obtained from the spectra in Fig.\,\ref{Fig02}a. The confined states and the lower edge of the continuum (indicated with a darker grey background) are shown with connected symbols. Horizontal lines identify the energy of the Brillouin laser and Stokes scattered photons. Double pointed black arrows highlight the situations where double optical resonance (DOR) is attained. 
\textbf{b} Symbols show the Brillouin intensity of the SL slow mode as a function of trap laser power, from the spectra in Fig.\,\ref{Fig02}b. Single (SOR) and double optical resonances are indicated with blue and red vertical arrows, respectively. The curve corresponds to the model discussed in the text.
}
\label{Fig03}
\end{figure}
 
 \paragraph*{\textbf{Single and double resonant cavity optomechanics.}} 
Figure.~\ref{Fig03}a presents a fan-plot with the energies of the 3D trap optical modes obtained from the spectra in Fig.\,\ref{Fig02}a. The energy of the BL and the Stokes photons scattered by the slow vibrational mode of the SL are also indicated. Figure.~\ref{Fig03}b  shows with solid symbols the Brillouin intensity associated to this slow mode, obtained from the spectra in Fig.\,\ref{Fig02}b. Starting from low TL powers, the two first blue vertical arrows highlight  situations in which SOR is attained. That is, the BL is resonant with a cavity mode (either the first or second confined modes), but the inelastically scattered Stokes photons are not in resonance with any mode. Note that we are in the side-band resolved regime \cite{Monroe,ReviewCOM,Kippenberg}, meaning that the frequency of the vibrational modes is significantly larger than the width of the optical cavity modes (or, equivalently,  the vibrational mode oscillates several periods before the trapped photon escapes from the resonator). This condition is critical for diverse cavity optomechanical phenomena, and particularly to attain efficient cooling (to be discussed below). In our experiments, one consequence of this side-band resolved condition is that the Brillouin signals for SOR are very weak (identified with blue vertical arrows in Fig.\,\ref{Fig03}).  This is particularly the case for excitation through the first confined trap state: with no available mode at lower energies the scattered photons cannot escape the resonator and thus the Stokes process is strongly inhibited. 

On increasing the TL power in Fig.\,\ref{Fig03}b the Brillouin signal strongly increases, displaying three clear maxima.  By correlating these maxima with the fan-plot in Fig.\,\ref{Fig03}a, it follows that the system successively goes through three DORs (indicated with red vertical arrows in the figure). These DORs involve first the excitation with the BL through the 3$^{rd}$ optical trap state and scattering through the 1$^{st}$ one ($3 \rightarrow 1$), and then with increasing power successively $4 \rightarrow 2$, and  $5 \rightarrow 3$. The magnitude of this trap-enhanced optomechanical coupling decreases as the order of the involved optical modes increases. As we will argue below, the  optomechanical coupling scales as $1/D$, where $D$ is the lateral diameter of the involved trap-confined mode \cite{AnguianoPRL,AnguianoScaling}. Because of the Gaussian shape of the laser-induced optical trap \cite{Anguiano}, modes are laterally less confined as their order increases. 
Summing up, the laser-induced emergence of discrete modes of the 3D optical trap leads to strongly enhanced optomechanical coupling. The accessed DORs could allow for optomechanical cooling and stimulation of the SL slow vibrational mode. This will be addressed next.

 \paragraph*{\textbf{Optomechanical cooling.}} 
Two of the relevant consequences of dynamical backaction in cavity optomechanics are the possibilities to cool, or alternatively stimulate, the mechanical motion by \textit{cw} laser excitation \cite{ReviewCOM}. The standard way to cool (stimulate) a mechanical mode, in the presence of a single optical mode coupled to this vibration, is by red-detuned (blue-detuned) laser excitation of the optical cavity mode. Typically this process is most efficient when the detuning between the laser and the involved optical cavity mode equals the energy of the vibrational mode that is being sought to cool or stimulate. While this method is the most standard and extended in the literature, it can become strongly weakened in the side-band resolved regime because the transmission of the pump photons into the resonator is strongly reduced (similar to what happens with the Stokes photons in the SOR discussed above). This problem can be overcome by using two optical cavity modes coupled by the mechanical motion \cite{Grudinin,Renninger,Otterstrom}, as we have shown by the DOR with the laser-generated optical traps in Fig.\,\ref{Fig03}b.

\begin{figure}[thhh!!!]
 \begin{center}
\includegraphics*[keepaspectratio=true, clip=true, trim = 0mm 0mm 0mm 0mm, angle=0, width=\columnwidth]{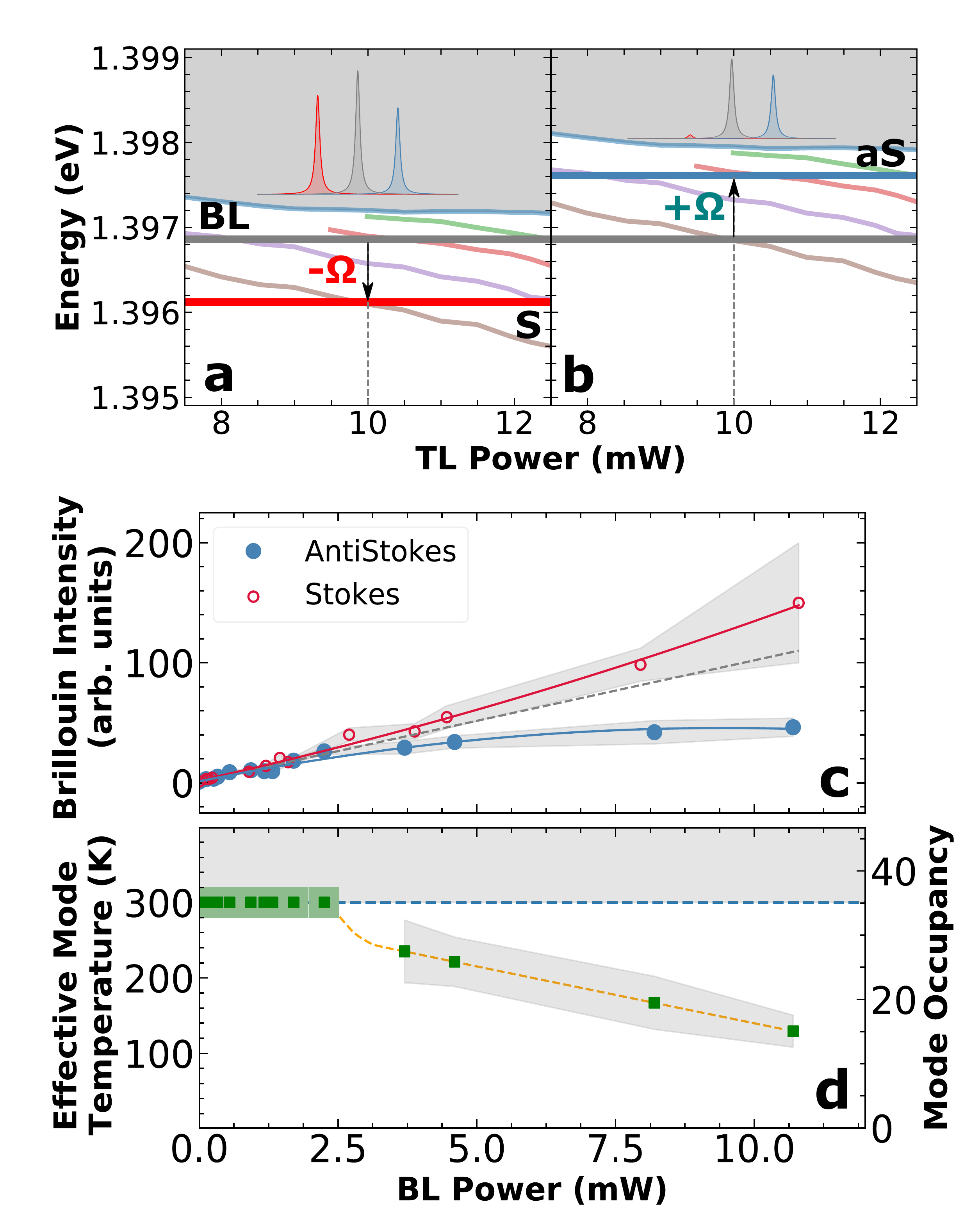}

\end{center}
\caption{
\textbf{Optomechanical cooling with a laser induced optical trap.}
\textbf{a} Schemes of the Stokes (stimulation), and \textbf{b} anti-Stokes (cooling) geometries. The coloured curves are the fan-plots describing the optical trap modes as in Fig.~\ref{Fig03}a. The energy of the Brillouin laser (BL), and the Stokes (S) and anti-Stokes (aS) photons are indicated with horizontal lines.  $\pm \Omega$ corresponds to the energy of the SL slow  vibration. Note that in \textbf{a} both the Stokes and the anti-Stokes channels are allowed with available optical modes. In \textbf{b} the Stokes channel is strongly inhibited with no optical modes accessible at energies lower than that of the BL. The expected Brillouin spectra resulting from these available resonances are schematized at the top of \textbf{a} and \textbf{b}.
\textbf{c} Experimental intensity of the S and aS signals as a function of BL power, obtained for the resonance conditions depicted in \textbf{a}  and \textbf{b} , respectively, and a TL power of 10~mW (DOR condition in Fig.~\ref{Fig03}). The dashed straight line is a linear fit based on the low power data. Coloured regions indicate the experimental uncertainty. The red and blue curves are guides to the eye. 
\textbf{d} Effective mode temperature (or equivalently mode occupation) for the slow SL vibrational mode deduced from the anti-Stokes intensities in \textbf{c}. The yellow dashed curve is a guide-to-the-eye.
}
\label{Fig04}
\end{figure}

Based on the previous considerations we chose a two-mode scheme as depicted in Figs.\,\ref{Fig04}a and b. The TL power is set at the condition of the strongest DOR shown in Fig.\,\ref{Fig03}b ($P_\mathrm{TL}\sim$10\,mW). Without changing the BL energy, the system is modified to operate in cooling (anti-Stokes) or stimulation (Stokes) conditions simply by displacing the spot on the structure and thus rigidly shifting the optical cavity modes. In this way, for the two situations the same two optical trap modes participate, but interchanging their role. Moreover, the detuning between the BL and the SL exciton resonance is constant, so that for the two processes the laser exciton-mediated photoelastic resonance is the same.~\cite{Jusserand} This is done in addition to exclude any possible effect of differential absorption and eventually heating associated with the BL laser. Thus, the main difference between the intensity of the Stokes and anti-Stokes processes is the intervening boson factor, which is proportional to $n+\mathrm{1}$ for the creation of a phonon (Stokes) and to $n$ for a destruction of a phonon (anti-Stokes), with $n$ the phonon occupation.  In this way the Stokes and anti-Stokes intensities as a function of laser power can be used to obtain $n$ and thus probe the presence of either vibrational pumping or cooling \cite{Kneipp,Maher,Harris}.

Figure~\ref{Fig04}c presents the BL power dependence of the Stokes (red) and anti-Stokes (blue) signal intensities. A linear dependence with BL power is expected in spontaneous Brillouin scattering (indicated with a dashed straight line extrapolated from the region of low power). Departures from this linear dependence are a signature of higher-order processes (that is, that $n$ is not determined by thermal equilibrium, but becomes dependent on the BL power). The Stokes contribution follows quite closely a linear dependence, with a possible weak supralinear behavior above $\sim$3\,mW. The anti-Stokes signal however clearly departs from linearity around  $\sim$2.5\,mW, displaying a systematic sub-linear behavior. From this latter departure, and the fact that at low BL powers the system is in thermal equilibrium (linear region), we can extract the phonon population as a function of BL power for the anti-Stokes configuration, or equivalently the effective temperature of the mode (see Fig.\,\ref{Fig04}d). A significant mode cooling from room temperature down to $\sim$120\,K is demonstrated, reaching a mode occupation of only $\sim$14 without cryogenic refrigeration. This is the second central result of this work. We note that the anti-Stokes scheme depicted in Fig.\,\ref{Fig04}b is ideal for optomechanical cooling due, firstly, to the existent DOR mediated by 3D optical trap modes, and secondly,  because of the side-band resolved regime with no modes available at energies lower than that of the BL, thus strongly inhibiting the competing Stokes processes \cite{Monroe,ReviewCOM,Kippenberg}. For the Stokes geometry, this is not the case:  optical modes present at energies higher than that of the BL contribute through anti-Stokes channels to balance the phonon generation, and thus to limit the possibility to access a self-oscillation threshold with the used Gaussian photon potential.

 \paragraph*{\textbf{Optomechanical coupling and cooperativity.}} 
 
 The main mechanism leading to the optomechanical coupling in SL-embedded semiconductor microcavities and close to excitonic resonance is photoelastic (i.e., related to an electrostrictive optical force) \cite{QW-comb}. The magnitude of the single-photon photoelastic coupling rate $g_\mathrm{0}$ can be calculated from the overlap integral of the normalized mode profiles of the incident and scattered optical [$\mathcal{E}(z)$] and strain [$\partial_zu_{\textrm{m}}(z)$] fields as \cite{Renninger}:
\begin{eqnarray}
g_0=\mathcal{K} \mbox{$\frac{1}{\sqrt{D_{\textrm{i}}\,D_{\textrm{s}}}}$}\!\int_{L} \!\epsilon_{\textrm{r}}^2(z)\,p_{12}(z)\partial_zu_{\textrm{m}}(z)\,\mathcal{E}^{\ast}_{\omega_{\textrm{s}}}(z)\,\mathcal{E}^{}_{\omega_{\textrm{i}}}(z)\,dz,\quad
\label{Eq}
\end{eqnarray}
where $\mathcal{K}=\frac{1}{\epsilon_{\textrm{r}}^{\text{eff}}\,L_{\textrm{opt}}^{\text{eff}}}\sqrt{\frac{\hbar\,\omega_{\textrm{s}}\,\omega_{\textrm{i}}}{2\,\pi\,\Omega_{\textrm{m}}}}$. 
Here $\epsilon_{\textrm{r}}^{\text{eff}}=d_{\textrm{T}}^{-1}\sum_j \epsilon_j\,d_j$, $d^{}_{\textrm{T}}$ is the total structure's thickness and $d_j$ the width of each individual layer of the heterostructure.
$L_{\textrm{opt}}^{\text{\tiny eff}}$ is the spacer thickness plus the contribution of the exponential penetration of the field into the DBRs. $L$ is the samples thickness. $\epsilon_\mathrm{r}$ is the media relative permittivity. 
$p_{12}$ is the material-dependent photoelastic constant, resonant in the GaAs quantum wells. And $\omega_{\textrm{i}}$($\omega_{\textrm{s}}$) and $D_{\textrm{i}}$($D_{\textrm{s}}$) are the angular frequency and effective lateral diameter of the incoming(scattered) photon mode, respectively. This expression highlights the relevance of having: i) a good overlap between light and strain fields (something attained for the SL slow vibrational mode, see Supplementary Note 3), ii) a large photoelastic coupling (existent in GaAs QWs, as described in Ref.\,[\onlinecite{Jusserand}]), and iii) a small optical mode diameter, as accomplished through the Gaussian laser engineered optical trap.
Based on published material parameters and structural information of our resonator, with the photon and vibrational modes of the structure evaluated with standard methods applicable in layered media (see the Supplementary Note 3 for details), we compute $g_\mathrm{0}/2\pi=1.7$\,MHz for the SL slow mode at $\Omega_{m}=2\pi\times180$\,GHz. This is a very strong optomechanical coupling, two orders of magnitude larger than the one expected from radiation pressure forces in these devices \cite{VillafaneOptoelectronicForcesPRB}.

The strength of the interaction in cavity optomechanics is quantified by the optomechanical cooperativity, $C=\frac{4g_\mathrm{0}^2n_\mathrm{cav}}{\kappa \Gamma_\mathrm{m}}$. Here, 
$n_\mathrm{cav}$ is the number of photons in the cavity, and $\kappa$  and $\Gamma_\mathrm{m}$ are the photon and mechanical decay rates, respectively \cite{ReviewCOM}. The coupled optomechanical equations in the case of \textit{two} optical cavity modes precisely detuned by the frequency of the mechanical mode, lead to an optomechanically modified phonon effective lifetime $\Gamma_\mathrm{eff}=\Gamma_\mathrm{m} (1 \pm C)$ \cite{ReviewCOM,Renninger}. Here the minus sign corresponds to stimulation (the laser exciting in the upper energy mode), while the plus sign corresponds to cooling (excitation is done on the lower energy mode).  Non-linearities related to cooling or self-oscillation require that the cooperativity $C$ becomes of the order or larger than 1.   For example, for higher mode excitation if $C \geq1$ the system undergoes a transition to self-oscillation. The power required to attain $C=1$  ($\Gamma_\mathrm{eff}=0$) defines the phonon ``lasing'' threshold condition.

The self-oscillation threshold power for the two-mode ($P^{(2)}_\mathrm{Th} $) configuration is related to that for one-mode threshold ($P^\mathrm{(1)}_\mathrm{Th} $) by $P^\mathrm{(2)}_\mathrm{Th} = P^\mathrm{(1)}_\mathrm{Th}/\left(1+4\frac{\Omega_\mathrm{m}^2}{\kappa^2}\right)$.\cite{Renninger} This expression highlights the relevance of working with two modes in the strong side-band resolved regime. Indeed, for the system under consideration $\Omega_\mathrm{m}/2\pi \sim 180$~GHz$>\kappa/ 2\pi \sim 75$~GHz). That is, $P^{(1)}_\mathrm{Th} \sim 24 \times P^{(2)}_\mathrm{Th}$. The factor $\times 24$ arises due to the fact that one of the two photons intervening in the optomechanical process is detuned respect to the optical cavity mode and thus a larger external laser power is required to inject the intra-cavity photons required to attain self-oscillation. This explains the smaller intensity of the SORs, when compared to the DOR processes, in Fig.~\ref{Fig03}b. Consideration of the relative contribution of single and double resonant optical processes based on this factor, and that of the mode lateral size-dependence $1/D$ discussed above and extracted from the experimental spatial images as in Fig.~\ref{Fig01}, allows to phenomenologically model the Brillouin intensity dependence with TL power as shown with a continuous line in Fig.\,\ref{Fig03}b (details of this procedure are included in the Supplementary Note 4).

While this model well describes quantitatively  the relative contribution of the different resonances, evaluation of the conditions needed for cooling or self-oscillation, requires additional considerations. In fact, we note that there are important assumptions implicit in the equation $\Gamma_\mathrm{eff}=\Gamma_\mathrm{m} (1 \pm C)$, the most relevant being that only the two described DOR optical modes are available for the optomechanical processes, with no competing inverse channels present. It turns out that this is a reasonable description of the anti-Stokes geometry in Fig.\,\ref{Fig04}b but, as we have discussed above, not for the Stokes one in Fig.\,\ref{Fig04}a. We thus concentrate in what follows only  in the cooling geometry.  
With $g_\mathrm{0}$ obtained from Eq.\,\eqref{Eq} above, we can evaluate the cooperativity $C$. Using $\kappa=2\pi\times 75$~GHz,  $\Gamma_m=2\pi\times160$~MHz,  and the intracavity photon number $n_\mathrm{cav}=P/(\hbar\omega\kappa) \approx 2 \times 10^{6}$  for a BL power of $P_\mathrm{BL} \sim10$~mW as used in the cooling experiment of Fig.\,\ref{Fig04}b,  we obtain $C \sim 2$.  Thus, a reduction of the phonon occupation by a factor of $\sim 3$ is expected from this estimation ($\Gamma_\mathrm{eff} \sim 3~\Gamma_\mathrm{m}$), a value that is notably coincident with the experiments. Note also that this model for  $\Gamma_\mathrm{eff}$ is consistent with a reduction of the mode occupation that is linear with the BL power (i.e., linear with $n_\mathrm{cav}$), something also consistent with the experimental findings shown in Fig.\,\ref{Fig04}d.

\section{Discussion} 
 
We have demonstrated that laser-generated 3D optical traps  in a planar semiconductor microcavity lead to a full discretization of the photon spectra, and through it to a strong enhancement of the optomechanical coupling by single and double optical resonances. As compared to laser trapping using etched pillars or laterally structured spacers of otherwise planar microcavities, the proposed scheme has several advantages. Namely, i) it does not require technologically complex techniques of device microfabrication, ii) the laser-engineered potentials can be dynamically modified on demand, and iii) microstructured lateral edges are avoided, which are known to be limiting factors for both the photon and phonon lifetimes for microstructures with lateral size below a few microns.\cite{AnguianoPRL,AnguianoScaling} Disadvantages are the requirement of a laser for the establishment of the lateral potential, and the eventual related heating associated to the latter.

In our experiments a laser-generated Gaussian trap was used for the optomechanical cooling of the slow $\sim$180\,GHz vibrational mode of a GaAs/AlAs superlattice by $\sim$200\,K from room temperature down to $\sim$120\,K (mode occupation of only $\sim$14) without cryogenic refrigeration. While the reported Gaussian optical trap is not appropriate to access non-linearities in Stokes processes (self-oscillation), the large magnitude of the electrostrictive optomechanical coupling and the strong side-band resolved situation accessible in these semiconductor devices, imply that proper design of the photon potential, to quench the limiting anti-Stokes channels, should allow for phonon lasing.  It is straightforward to extend the results reported here to the polariton regime, for which both attractive and repulsive laser-controlled effective potentials can be designed. In this way, our results pave the way for  reconfigurable cavity optomechanics applications based on laser potential landscape engineering, opening new roads in the conception of hybrid devices based on cavity quantum electrodynamical and cavity optomechanical phenomena \cite{NatComm}.

\section{Methods}
\paragraph*{\textbf{Sample description.}} 
The experiments were performed at room temperature in a \textit{planar} microcavity  grown on a (001)-oriented GaAs substrate by molecular-beam epitaxy. It consists of two distributed Bragg reflectors (DBRs) enclosing a thick resonant spacer with a large optical path length of $9\lambda/2$. The top (bottom) optical DBR is formed by 15.5 (18.5) periods of Al$_{0.1}$Ga$_{0.9}$As/Al$_{0.95}$Ga$_{0.05}$As bilayers optimized to confine an optical mode with a quality factor $Q_{\textrm{opt}} \sim 4.5 \times 10^3$. The optical spacer of the cavity is composed of a superlattice (SL) formed by 41.5 periods of  17.1\,nm/7.9\,nm GaAs/AlAs bilayers. This embedded SL is identical to the one studied in Ref.\,[\onlinecite{Jusserand}] to demonstrate ultrastrong optomechanical coupling based on polariton resonances. In our experiments, we exploit this resonant behavior for an enhanced optomechanical signal but remain always well below the exciton gap ($\sim$1.44\,eV at room temperature), so that the cavity mode is almost $100\%$ photonic in character. The SL is designed to confine slow acoustic phonons \cite{TrigoSurfaceAvoding,TrigoPRL,Villafane} at $\sim$180\,GHz ($\sim$5.7\,cm$^{-1}$). This Brillouin zone-center mode has almost zero group-velocity because it is located at the lower edge of the first phononic bandgap of the SL. This slow speed leads to an effective mechanical quality factor $Q_m \sim 10^3$.  The sample is tapered so that the optical modes energies can be varied by displacing the laser spot position. \\

\paragraph*{\textbf{Experimental details.}} 
A custom-made microscope configuration is used to focus three lasers on perfectly overlapping $\sim$4\,$\mu$m spots. Firstly, the 514.5\,nm green line of an Ar-Kr laser (``trap laser'', TL) is used to locally heat the structure and, through the change of index of refraction with temperature \cite{Talghader}, generate a three-dimensional optical resonator (the planar cavity confines light along the axis of the heterostructure, while the Gaussian spot defines the lateral in-plane trap) \cite{Anguiano}. Because this wavelength is above the gap of the two materials forming the DBR layers, it is fully absorbed in the top DBR. This was set to avoid the excitation of electron-hole pairs from the SL by the TL. Photoluminescence  is typically overwhelmingly larger than the inelastic scattering contributions, so that this contribution would mask the optomechanical signals we want to detect. Secondly, and to excite the SL photoluminescence when required to better identify the spectral position of the generated three-dimensional optical modes, we use a \textit{cw} Titanium-Sapphire laser. This laser is  tuned to the DBR edge-modes to efficiently access the SL without further heating the structure, and above the SL exciton gap in order to locally excite its photoluminescence. This ``photoluminescence laser'' (PLL) is used with very low power (typically some $\mu$W) once the mW power of the TL (and thus the trapping potential) is chosen. It is used to characterize the optical modes of the 3D cavity, but is blocked prior to the Brillouin experiments. Finally, an additional single-mode stabilized Titanium-Sapphire ring-laser tuned to the 3D optical trap modes is used to resonantly excite the SL vibrational modes (``Brillouin laser'', BL). The power of this laser is always kept below a few mW. Its energy (typically $\sim$1.39\,eV) is well below the gaps of the DBRs and SL materials, and thus its contribution to additional heating is negligible. Although zero-wavevector phonons, as the studied slow SL mode, would require in principle a forward-scattering geometry to be coupled by inelastic light scattering, they are readily accessed when exciting through an optical cavity mode in a back-scattering configuration \cite{FainsteinPRL1995,TrigoFiniteSize}. \\

\section*{Data availability}
The source data that support the findings of this study are available from the corresponding author upon reasonable request. All these data are directly shown in the corresponding figures without further processing.



\section*{Acknowledgements} We acknowledge partial financial support from the ANPCyT-FONCyT (Argentina) under grants PICT2015-1063 and PICT-2018-03255, and from the U.N.Cuyo (Argentina) under grants SIIP C034 and 06/C555.  

\section*{Author contributions}
All authors contributed to all aspects of this work.

\section*{Competing interests}
The authors declare no competing interests.

\section*{Additional information}


\paragraph*{\textbf{Correspondence}} and requests for materials should be addressed to A.F.

\onecolumngrid

\pagebreak

\setcounter{section}{0}
\setcounter{page}{1}
\setcounter{figure}{0}
\renewcommand{\thesection}{Supplementary Note \arabic{section}}
\renewcommand{\figurename}{\textbf{Supplementary Figure} \!\!}
\renewcommand{\thefigure}{\bf\arabic{figure}}

\begin{center}
\textbf{\Large Supplementary Information for:\\~\\ \large Cavity Optomechanics with a Laser Engineered Optical Trap}
\end{center}

\section{Sample details}\label{sec: Sample details}

The sample studied corresponds to a \textit{planar} semiconductor microcavity. It was grown on a (001)-oriented GaAs substrate by molecular-beam epitaxy. The cavity consists of two distributed Bragg reflectors (DBRs) enclosing a thick resonant spacer with a large optical path length of $9\lambda/2$, with multiple quantum wells (MQWs) embedded, as schematized in the Supplementary Fig.\,\ref{FigS01}.
\begin{figure}[hhh!!!]
\begin{center}
\includegraphics*[keepaspectratio=true, clip=true, trim = 0mm 0mm 0mm 0mm, angle=0, width=0.7\columnwidth]{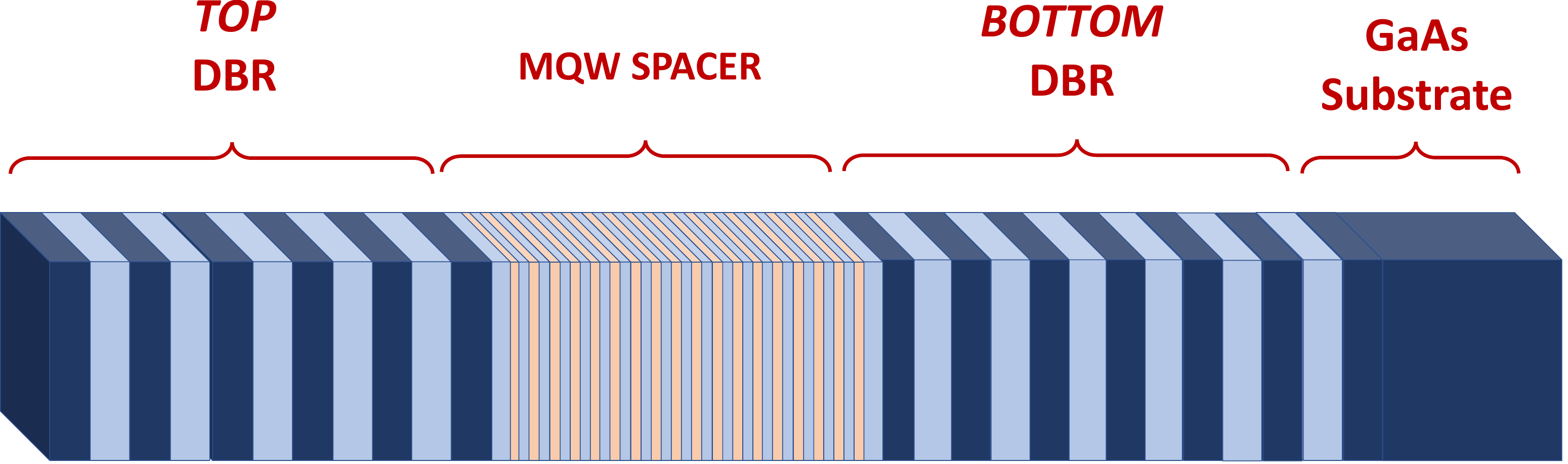}
\caption{Schematic representation of the studied sample.
The top (bottom) $\lambda/4$ DBR consists of 15.5 (19.5)  Al$_{0.1}$Ga$_{0.9}$As/Al$_{0.95}$Ga$_{0.05}$As periods, which constitute the \textit{mirrors} of the cavity, enclosing a $9\lambda/2$ optical spacer. The cavity spacer has 41.5 periods of GaAs/AlAs MQWs (see text for details).}
\label{FigS01}
\end{center}
\end{figure}
The cavity embedded MWQs consist of 41.5 periods of (17.1\,nm) GaAs/(7.9\,nm) AlAs, forming a superlattice (SL) that is meant to be identical to the one studied in Ref.\,[\onlinecite{Jusserand}], in order to demonstrate ultra-strong optomechanical coupling due to decoherence-avoidance in exciton-polariton resonances \cite{Rozas-PRB(2014)}. The whole nanostructure was grown tapered so that the optical cavity mode can be tuned in energy by varying the laser spot's position on the sample, around the excitonic energy $\sim 1.434\,$eV at room temperature (RT). 
\begin{figure}[t]
\begin{center}
\includegraphics*[keepaspectratio=true, clip=true, trim = 0mm 0mm 0mm 0mm, angle=0, width=0.5\columnwidth]{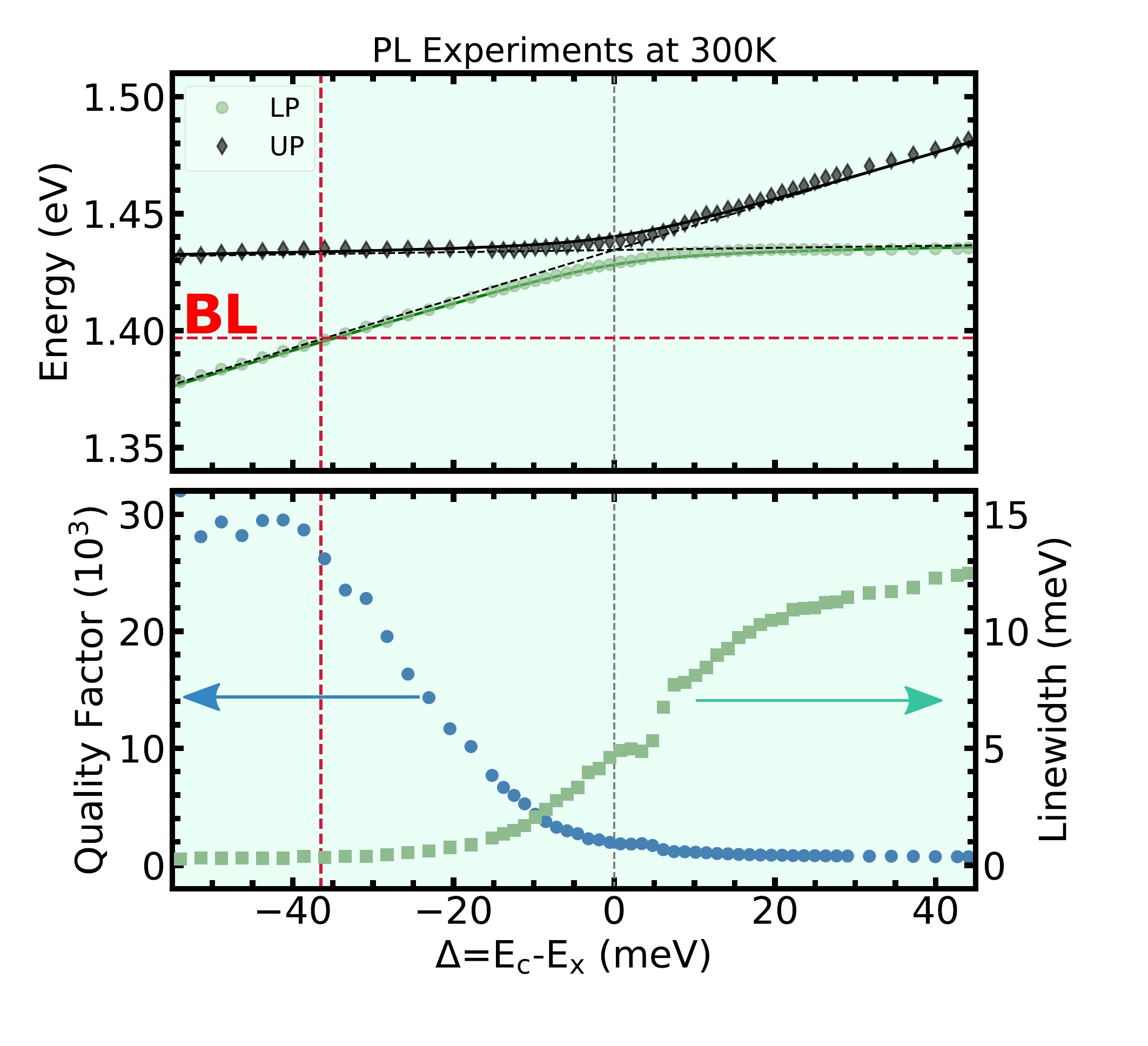}
\caption{
Top panel: PL peak's energy for cavity mode and MQW exciton as a function of the detuning $\Delta$, controlled by the laser spot position on the sample. LP and UP refer to Lower and Upper polariton resonances, respectively. The black dashed lines indicate the uncoupled photon-exciton system, that result from the simultaneous fitting (full lines) of the polariton branches with a simple two-coupled-state model. Bottom panel: Quality factor and FWHM for the LP resonance vs. $\Delta$. The red dashed lines indicate the Brillouin Laser energy and sample detuning set for the optomechanical experiments presented in this work.
}\label{FigS02}
\end{center}
\end{figure}
This is shown in Supplementary Fig.\,\ref{FigS02}. The top panel corresponds to the energy of the photoluminescence peak (measured at RT) as a function of the cavity-mode ($E_\textrm{C}$) and MQW-exciton ($E_\textrm{X}$) detuning, which is varied by changing the spot position on the sample. When the cavity mode is tuned close to the excitonic energy, the strong coupling regime is characterized by the well-known photon-exciton anti-crossing\cite{Weisbuch1992}, with a room temperature Rabi-splitting at $\Delta=0$\,meV of $E_\textrm{UP}-E_\textrm{LP}\sim$5\,meV. The two resulting cavity polariton branches, lower polariton (LP) and upper polariton (UP), change progressively their nature from photonic to excitonic (LP), and vice-versa from excitonic to photonic (UP), when passing through the anti-crossing.
For negative detuning, the LP initially behaves essentially as the cavity mode. When increasing the detuning, towards the anti-crossing, the quenching of the mode is observed with the consequent decrease of its quality factor. This is observed in the bottom panel of Supplementary Fig.\,\ref{FigS02}, where the linewidth (FWHM) of the LP's luminescence peak is plotted (light-green squares, right vertical axis). The left vertical axis of this panel corresponds to the evolution with the detuning of the quality factor (blue circles). All experiments are performed with the Brillouin laser (BL) tuned to $\sim$1.3968\,eV (corresponding to $\Delta = -37.7\,$meV), far from the zero-detuning condition, and exploiting an optical quality factor near to $\sim$3$\times 10^4$. The dashed red lines situate the position of the BL and the corresponding detuning.


\clearpage
\section{Dependence of the optical confined modes with the trap laser (TL) power}\label{sec: Dependence of the optical confined modes with the trap laser (TL) power}
 
In the Supplementary Fig.\,\ref{fig: PL dependence with power - image and spectrum} we show the photoluminescence (PL) acquired in spectral mode (bottom of each panel) and as color map recorded in calibrated spatial imaging mode (top of each panel). For each panel, the corresponding energy axis scale is the same. Panels (a) to (f) display the successive formation of higher confined modes for increasing power ($P_{\textrm{TL}}$), as indicated. As a reference, the laser used for Brillouin excitation (BL) is set attenuated at 1.3969\,eV and can be appreciated as a well-defined small peak or as a faint vertical line for the spectral or spatial map mode, respectively. The complete series displaying this evolution can be observed in the provided Supplemental Movie 1. 

\begin{figure}[h]
\begin{center}
	\begin{subfigure}{0.32\columnwidth} 
		\includegraphics[width=\textwidth]{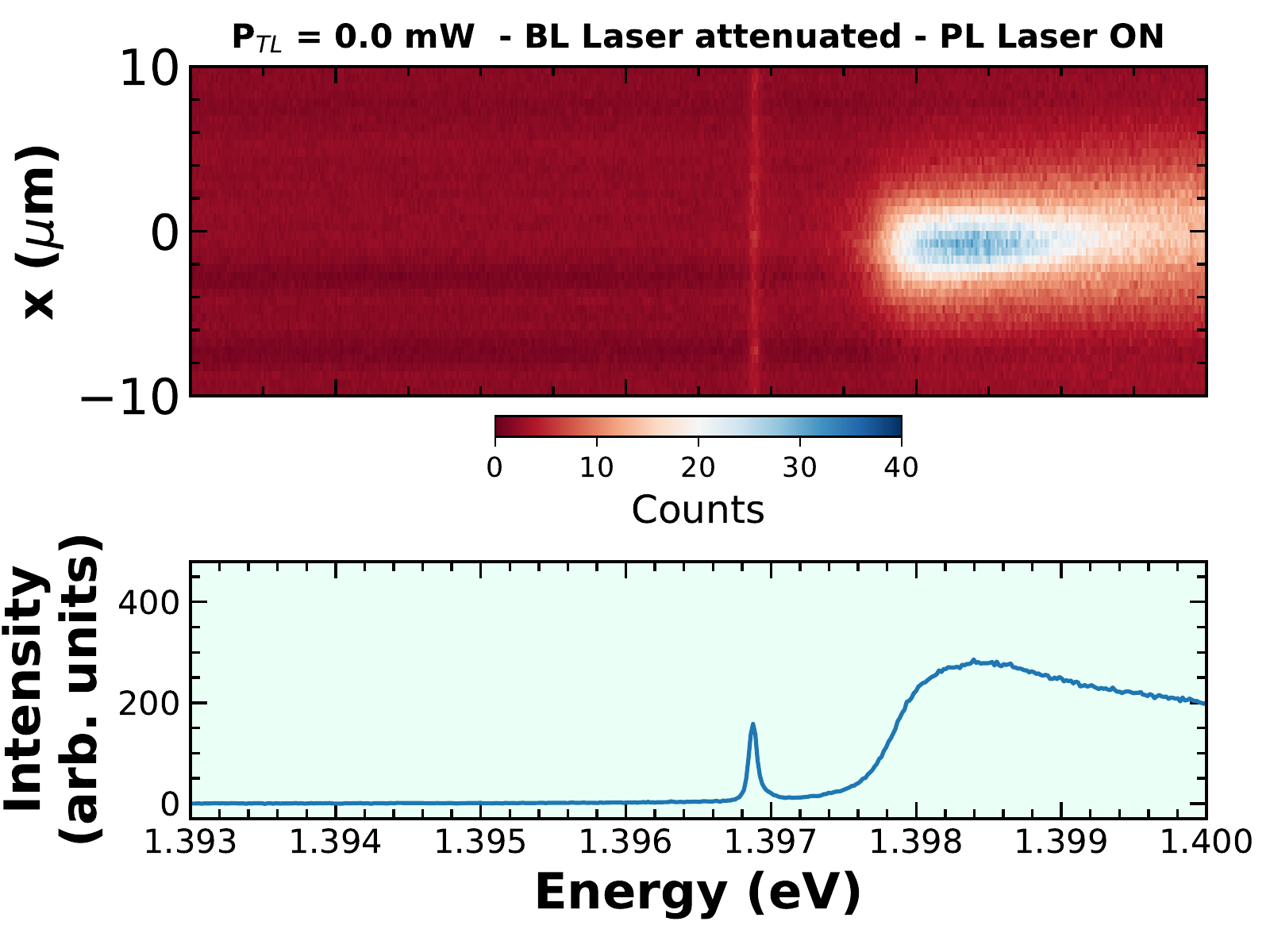}
		\caption{$P_{\textrm{TL}}=0.0$\,mW} 
	\end{subfigure}
	\begin{subfigure}{0.32\columnwidth} 
		\includegraphics[width=\textwidth]{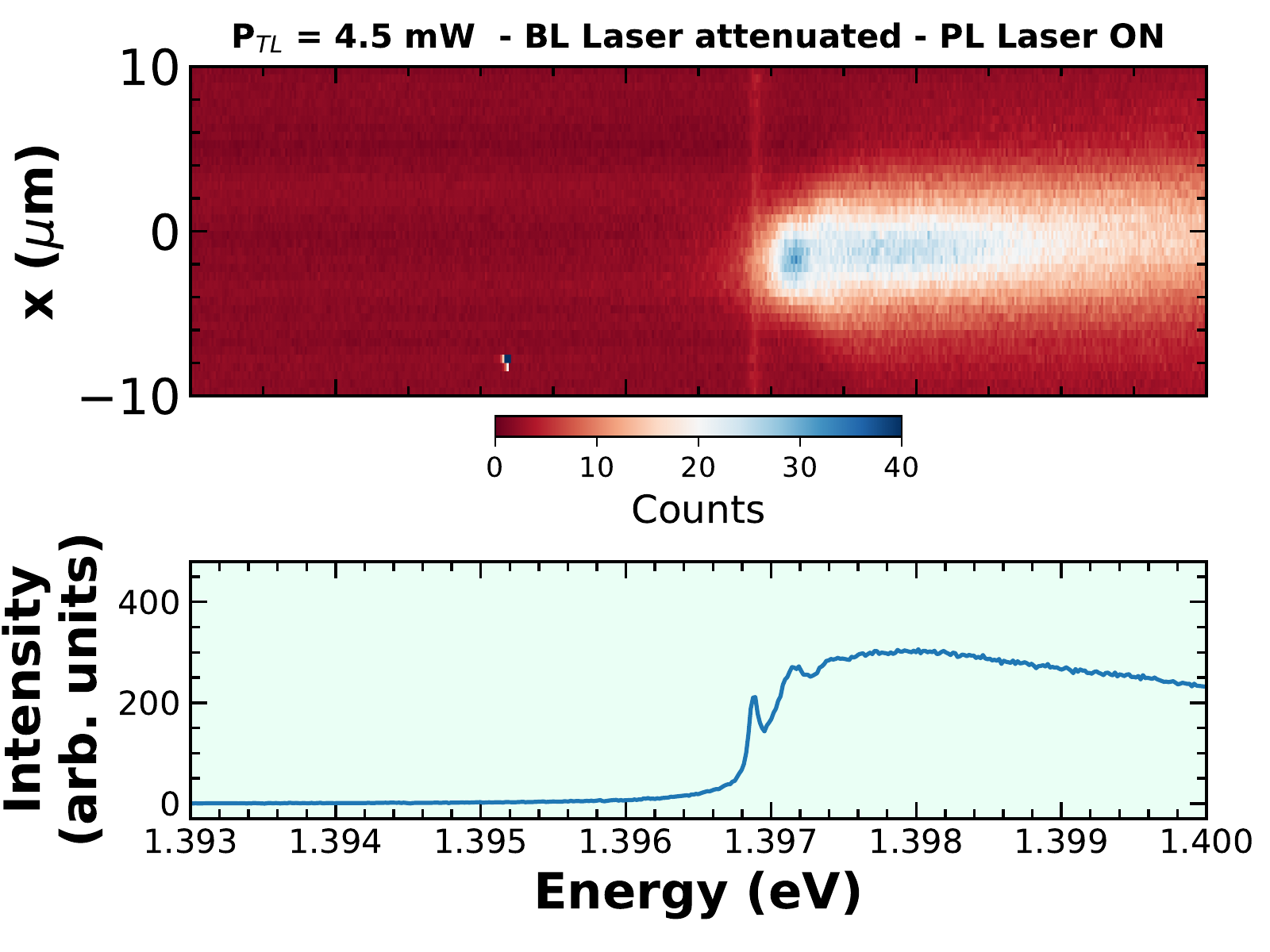}
		\caption{$P_{\textrm{TL}}=4.5$\,mW} 
	\end{subfigure}
	\begin{subfigure}{0.32\columnwidth} 
		\includegraphics[width=\textwidth]{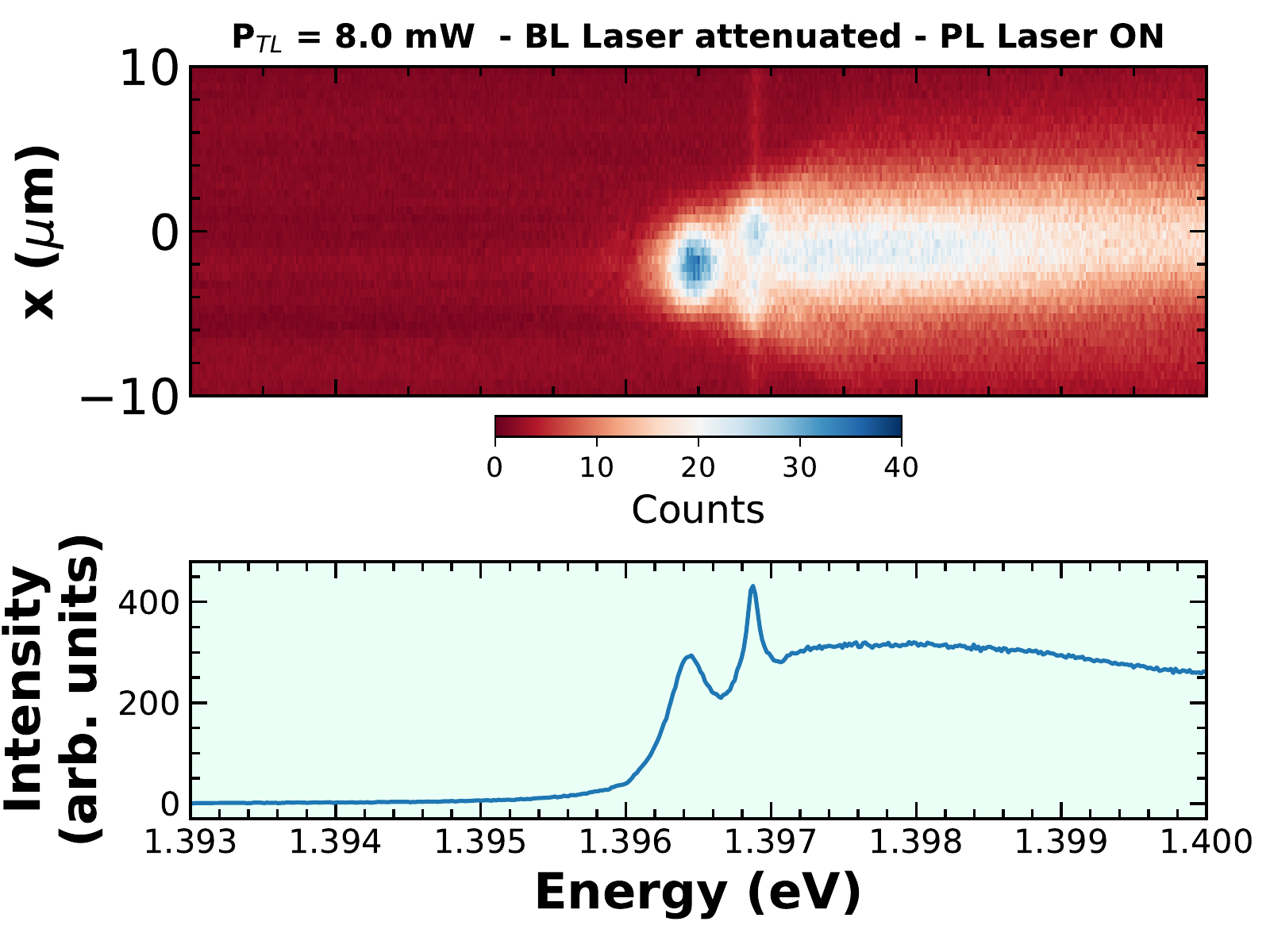}
		\caption{$P_{\textrm{TL}}=8.0$\,mW} 
	\end{subfigure}\\
	\begin{subfigure}{0.32\columnwidth} 
		\includegraphics[width=\textwidth]{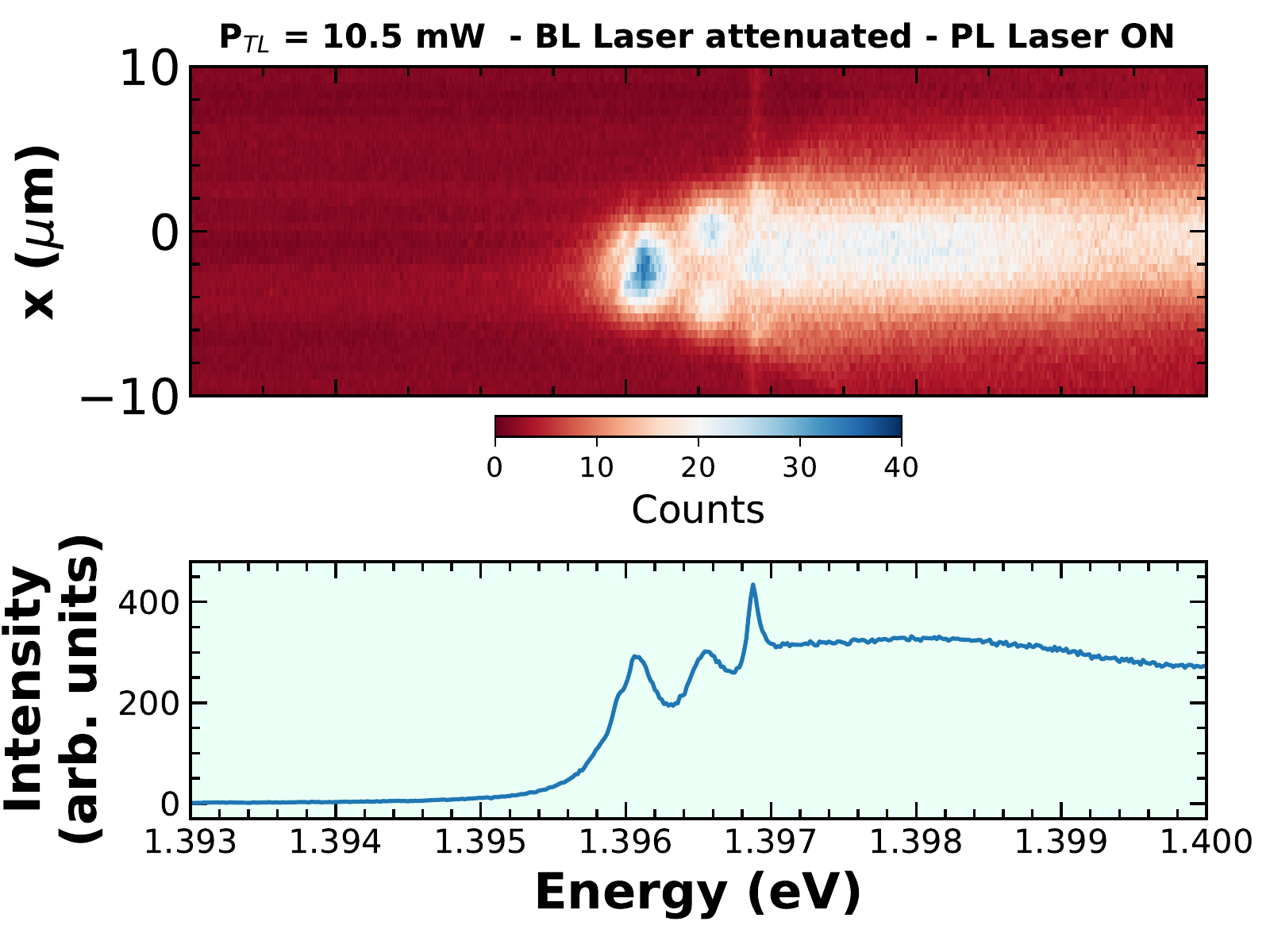}
		\caption{$P_{\textrm{TL}}=10.5$\,mW} 
	\end{subfigure}
	\begin{subfigure}{0.32\columnwidth} 
		\includegraphics[width=\textwidth]{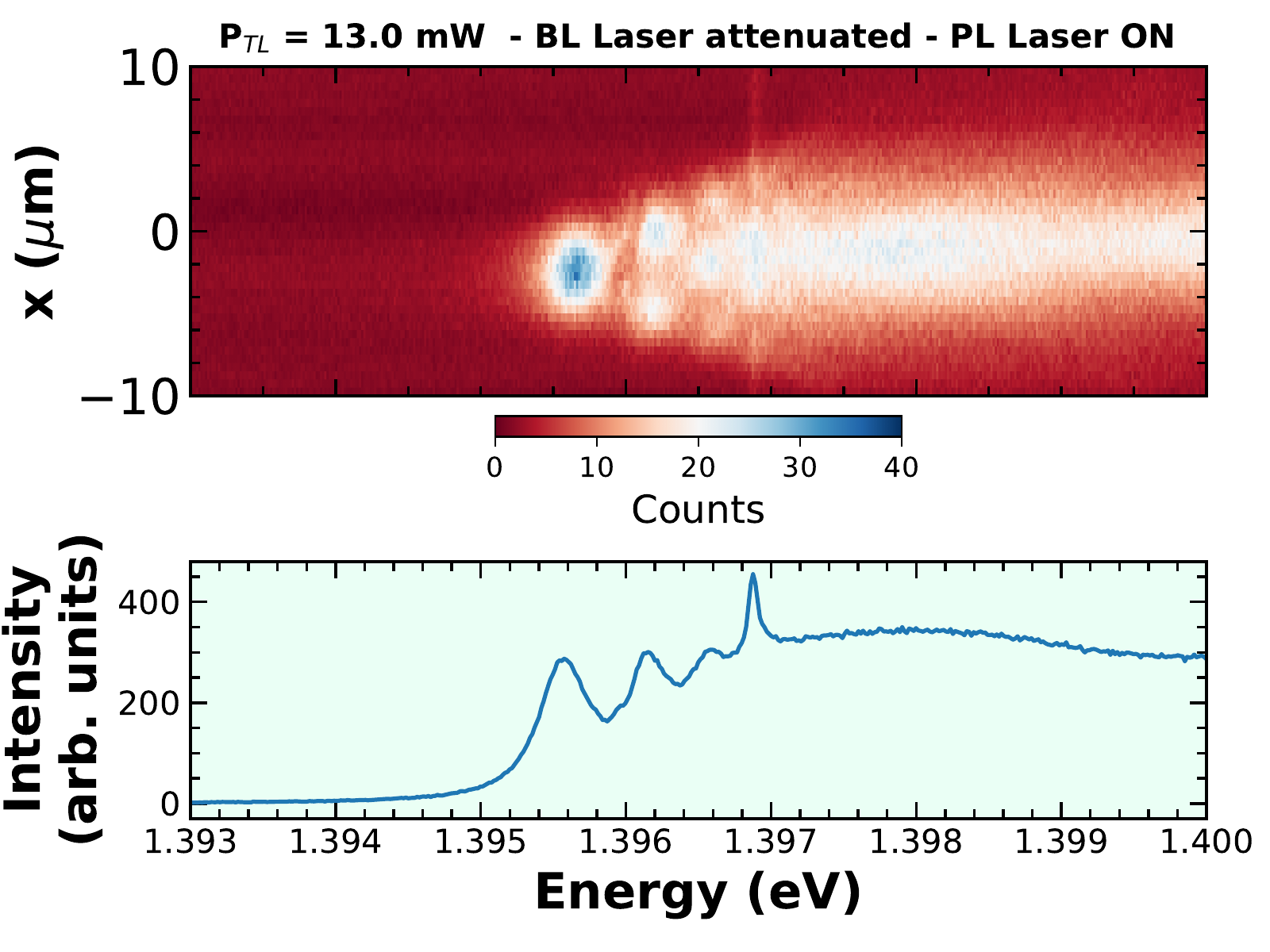}
		\caption{$P_{\textrm{TL}}=13.0$\,mW} 
	\end{subfigure}
	\begin{subfigure}{0.32\columnwidth} 
		\includegraphics[width=\textwidth]{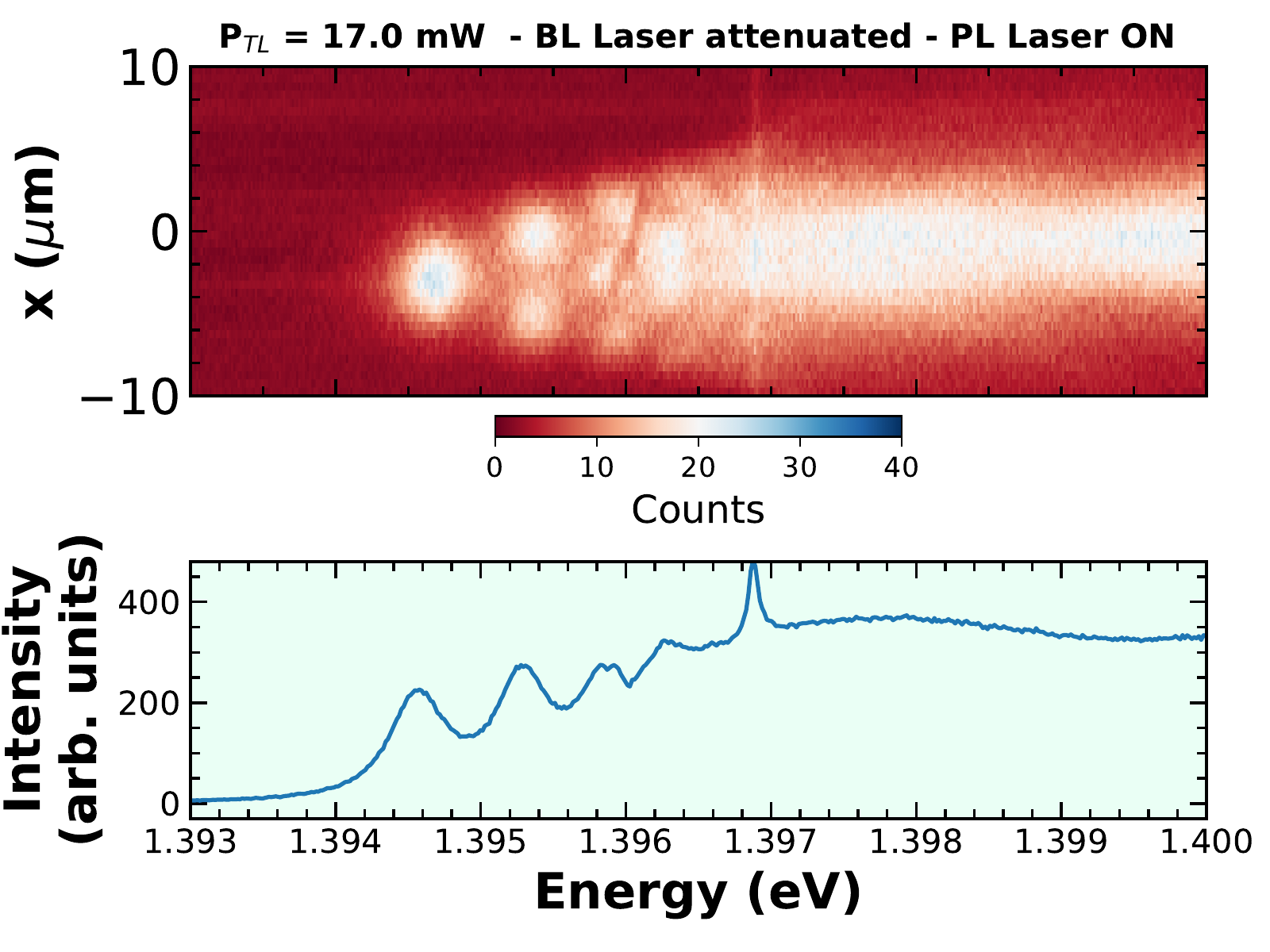}
		\caption{$P_{\textrm{TL}}=17.0$\,mW} 
	\end{subfigure}
	\caption{Photoluminescence (PL) calibrated spatial maps (top sub-panels) and corresponding PL spectra (bottom sub-panels) for varying trap laser power $P_{\textrm{TL}}$ as indicated in each panel from (a) to (f).} 
	\label{fig: PL dependence with power - image and spectrum}
\end{center}
\end{figure}

\subsection{Laser induced optical potential}\label{subsec: Laser induced optical potential}

Localized heating due to laser excitation by the TL gives rise to a refractive index gradient, which in turn allows for new optical modes to appear. These are spatially confined to the laser spot's area, and can be described quite well by a Gaussian potential well, as explained in Ref.[\onlinecite{Anguiano}]. Following the procedure introduced there, the parameters that describe the optical trap as well as the confined modes can be readily extracted. In Supplementary Fig.\,\ref{fig: 2a1} an example is presented for a TL power of $P_{\textrm{TL}}=$10\,mW, where the red dashed curve corresponds to the Gaussian trap that comes out from the fit \cite{Anguiano}. The depth of the trap ($\Delta \textrm{E}_0$), as well as the energy ($\textrm{E}_\textrm{fund}$) and effective width ($\textrm{D}_\textrm{eff}$) of the fundamental mode are indicated. 
This procedure is repeated for all TL powers where the confined modes are well distinguishable. This allows extracting the depth of the trap as a function of the TL power, which is presented in Supplementary Fig.\,\ref{fig: 2a2}, showing a rather linear behaviour. The complete set of measurements can be appreciated in the Supplementary Movie 2. \\
\begin{figure}[h]
\begin{center}
	\includegraphics[width=0.5\textwidth]{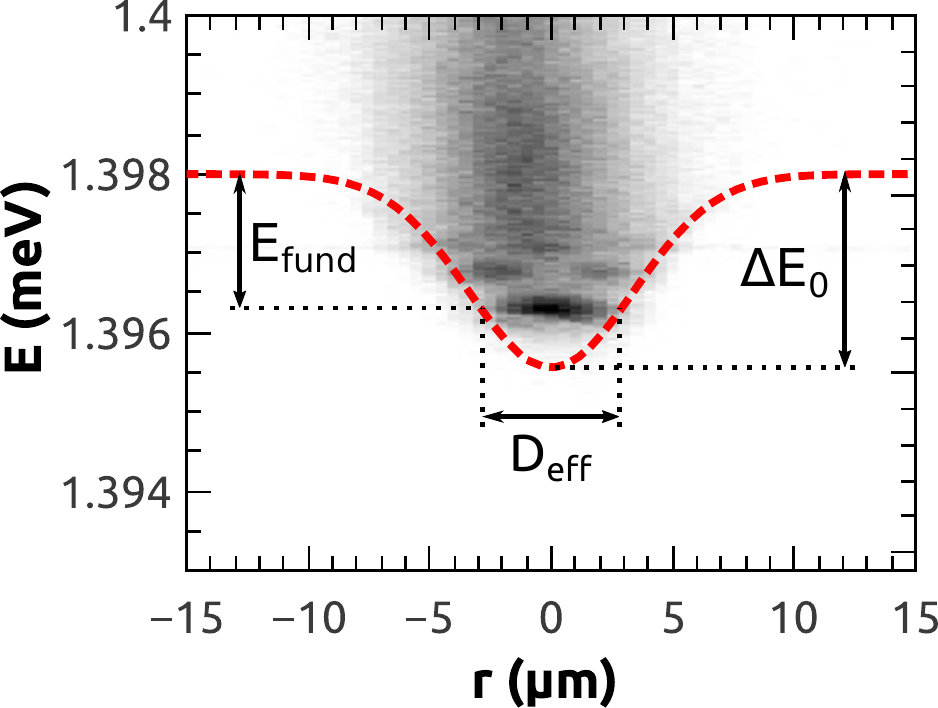}
	\caption{Calibrated PL spatial map measured for a TL power of $P_{\textrm{TL}}=$10\,mW. The red dashed line corresponds to the fitted Gaussian potential. $\Delta \text{E}_0$, $\text{E}_\text{fund}$ and $\text{D}_\text{eff}$ correspond to the trap depth, fundamental mode energy and effective width, respectively.}\label{fig: 2a1}

\end{center}
\end{figure}

\begin{figure}[h]
\begin{center}
	\includegraphics[width=0.5\textwidth]{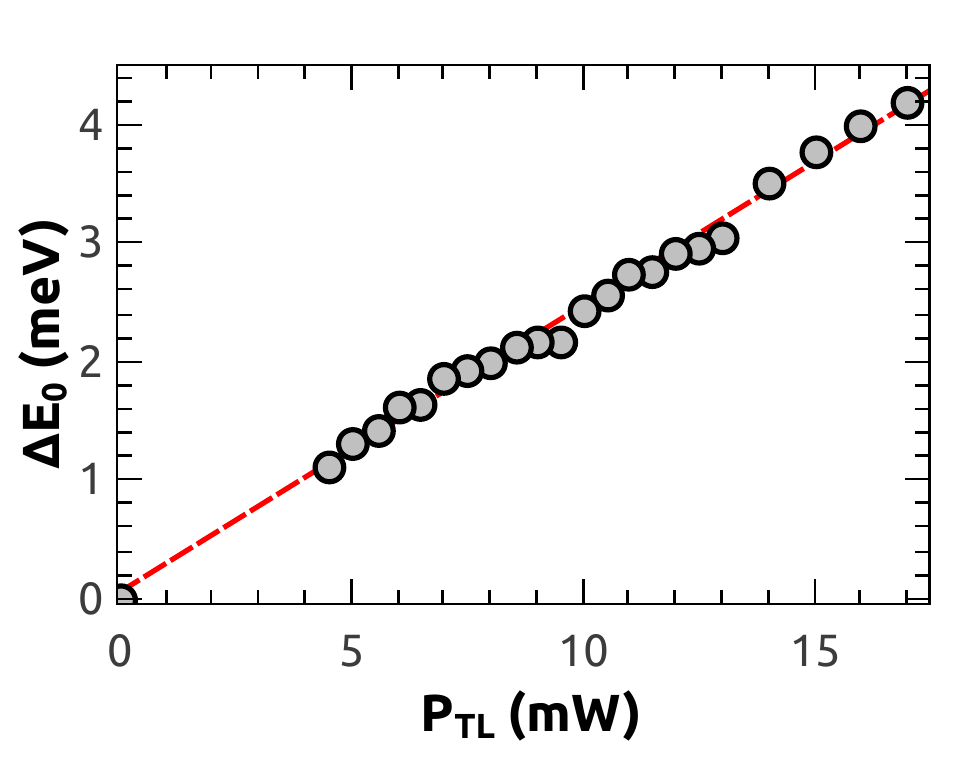}
	\caption{Power dependence of the laser-induced trap depth, as extracted through the method introduced in Ref.[\onlinecite{Anguiano}]. The red dashed line is a linear fit to the data.}\label{fig: 2a2}
\end{center}
\end{figure}

The effective width for each observed confined mode can be extracted (as exemplified in Supplementary Fig.\,\ref{fig: 2a1} for the fundamental mode) for each measured incident trap laser power. The result of this procedure is shown in Supplementary Fig.\,\ref{fig: 2a5}. Here a slight decrease of the effective with, i.e. the radial extension of the optical confined modes, for increasing trap laser power. The dashed vertical line indicated the  trap laser power $P_{\textrm{TL}}$ corresponding to the first DOR situation.\\
\begin{figure}[h]
\begin{center}
		\includegraphics[width=0.7\textwidth]{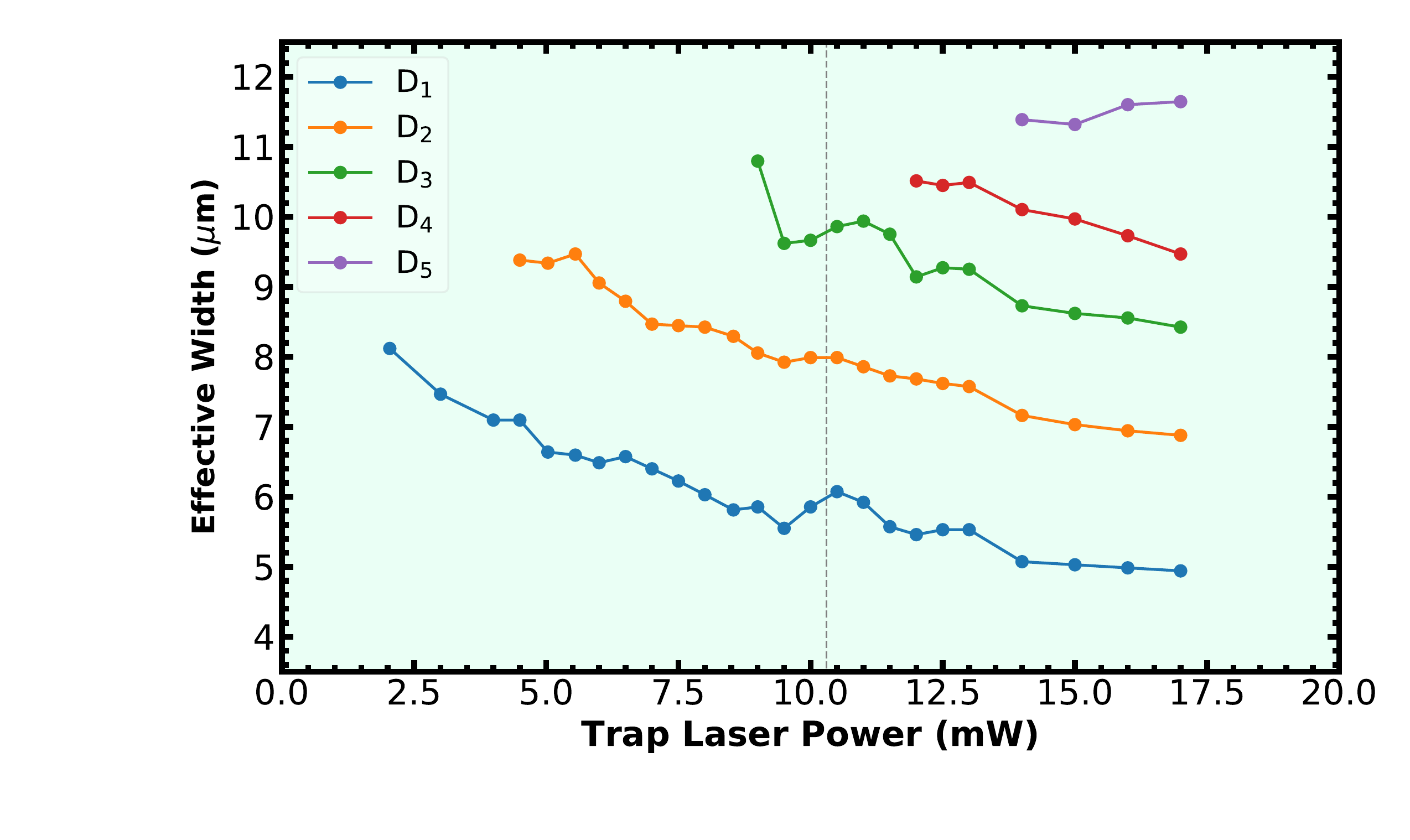}
	\caption{Effective width ($D_{\textrm{eff}}$) derived (as explained in the text) for each observed confined optical mode (as indicated), as function of the trap laser power $P_{\textrm{TL}}$. The vertical dashed line ($P_{\textrm{TL}}\sim$10\,mW) corresponds to the first double optical resonance (DOR).
	}\label{fig: 2a5}
\end{center}
\end{figure}

In order to further understand the physics involved in these laser-induced potential traps, we consider the fitted Gaussian model for the traps and solve the corresponding Schr\"odinger equation for the fundamental mode. To do so, a rotational in-plane symmetry was considered, and the finite difference Crank-Nicolson method was applied, with an imaginary time evolution. This is a relatively common way of finding the fundamental mode for an arbitrary potential, and does not require much in terms of computational power.\\

To be able to compare the result from the two different methods (the graphical method and the Schr\"odinger equation's solution), the obtained effective widths ($\text{D}_\text{eff}$) were normalized by the traps FWHM, and the obtained mode's energy ($\text{E}_\text{fund}$) was normalized by the obtained potential's depth ($\Delta \text{E}_0$). 
For the dimensionless solved Schrödinger equation, the energy had to be rescaled in order to fit the experimental values. This was performed for $\text{E}_\text{fund}$, which is shown in Supplementary Fig.\,\ref{fig: 2a3}. The same was applied for the effective width of the fundamental mode, and is depicted in Supplementary Fig.\,\ref{fig: 2a4}. Both show an excellent agreement. 
\begin{figure}[h]
\begin{center}
		\includegraphics[width=0.5\textwidth]{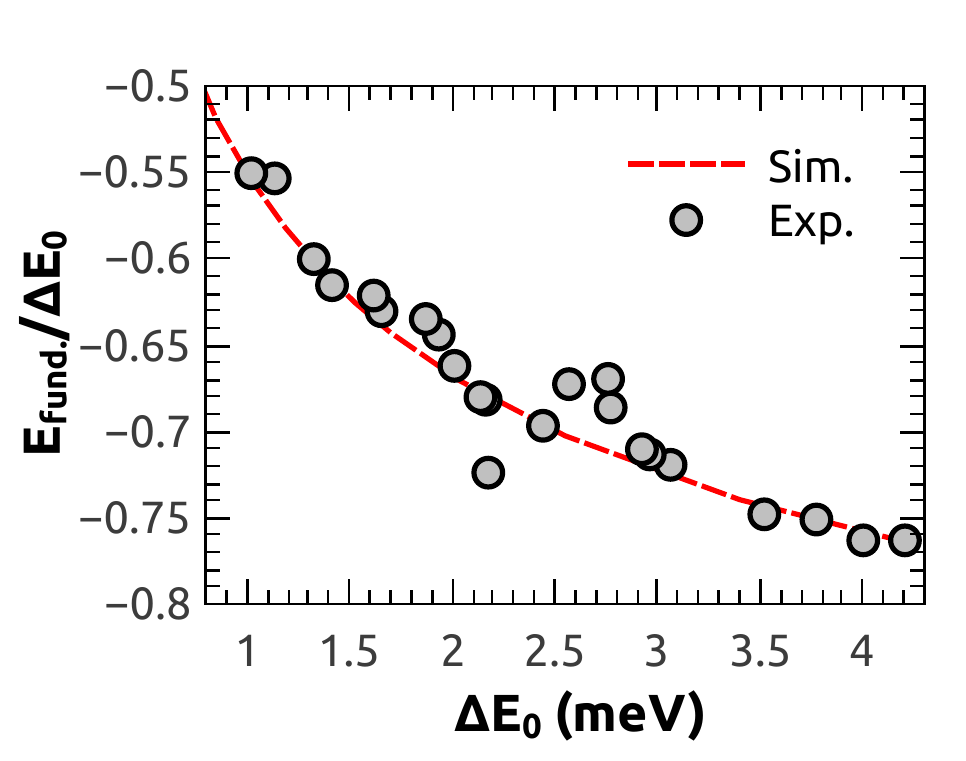}
	\caption{Fundamental mode's energy ($\text{E}_\text{fund}$) as extracted from the graphical method (symbols) introduced in Ref.[\onlinecite{Anguiano}], and by the numerical method (red dashed line). Both have been normalized by the traps' depth.}\label{fig: 2a3}
\end{center}
\end{figure}

\begin{figure}[h]
\begin{center}
		\includegraphics[width=0.5\textwidth]{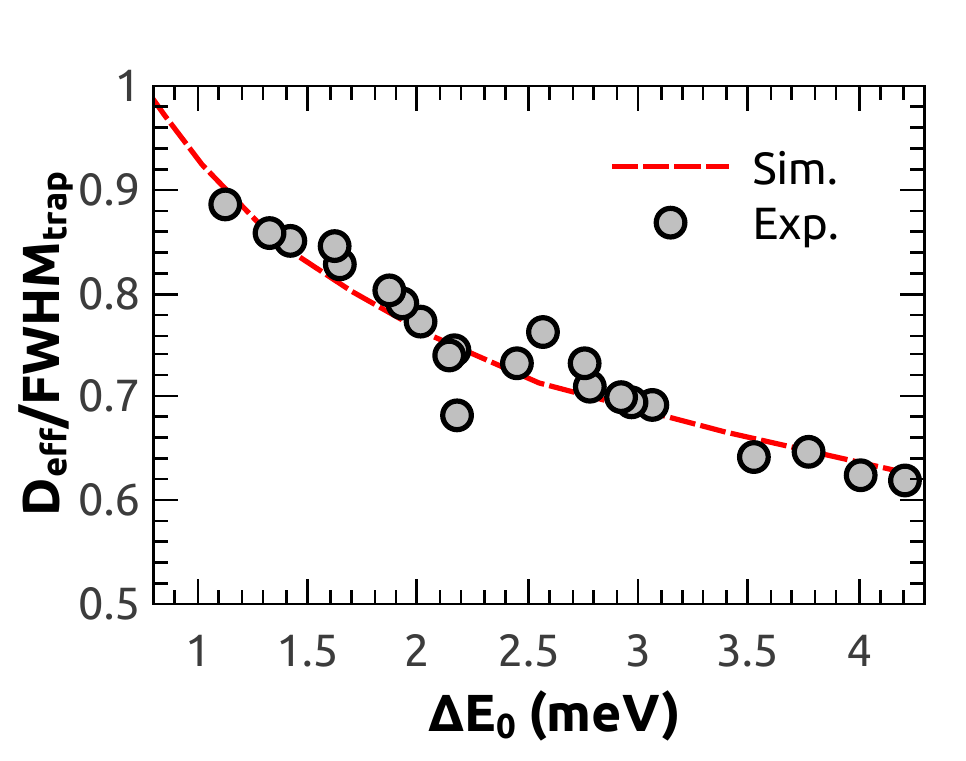}
	\caption{Fundamental mode's effective width ($\text{D}_\text{eff}$) as extracted from the graphical method (symbols) introduced in Ref.[\onlinecite{Anguiano}] and by the numerical method (red dashed line). Both have been normalized by the traps' FWHM.}\label{fig: 2a4}
\end{center}
\end{figure}

\clearpage

\section{Phonon modes of the system and Brillouin spectra}\label{sec: Raman spectra in the cavity - phonon peaks}
 
The vibrational eigenmodes of the heterostructure deserve a separate and more elaborate presentation. They are rather complicated, rising from the fact that the structure consists of a superlattice (SL) embedded in an optical microcavity that simultaneously acts as an acoustic cavity.

The vibrational acoustic modes are obtained using a 1D continuum elastic model based on a transfer matrix formalism [see \ref{sec: Single-photon photoelastic coupling rate} for details]. One way to understand the acoustic phononic structure of the system is to analyse the surface displacement as a function of the phonon energy. In Supplementary Fig.\,\ref{fig: acoustic displacement and Raman spectra} (top panels), plotted with the green curve, we present the surface displacement of a bare finite SL, i.e. the embedded SL of 41.5 periods but \textit{without} the optical microcavity. Superimposed, with the red curve, we show the tilted dispersion relation of an equivalent infinite SL showing the characteristic folding of the acoustic branch, and the opening of the characteristic acoustic mini-gaps\cite{Rytov, Jusserand-LightScattSol5(1989)}. These mini-gaps -phonon forbidden regions- are indicated by the gray-shaded areas. It is clear that for the finite SL (green curve) the surface displacement, at the mini-gap, is reduced to zero. The top-right panel corresponds to a zoom around the energy of interest ($\sim$5.7\,cm$^{-1}$). The violet curve corresponds to the calculated surface displacement for the complete structure, i.e. the SL embedded within the optical cavity. As is well observed, this curve behaves equivalently to the green curve (bare SL) at the gray mini-gap regions, which means that this contribution is inherited form the SL. Additional features (e.g. peaks) also arise, resulting from the acoustic contribution of the optical cavity spacer, as well as the contribution from the optical DBRs \cite{FainsteinPRL2013, Fainstein-LightScattSolIX(2007)}.

\begin{figure}[h]
\begin{center}
\includegraphics*[keepaspectratio=true, clip=true, trim = 0mm 0mm 0mm 0mm, angle=0, width=1.\columnwidth]{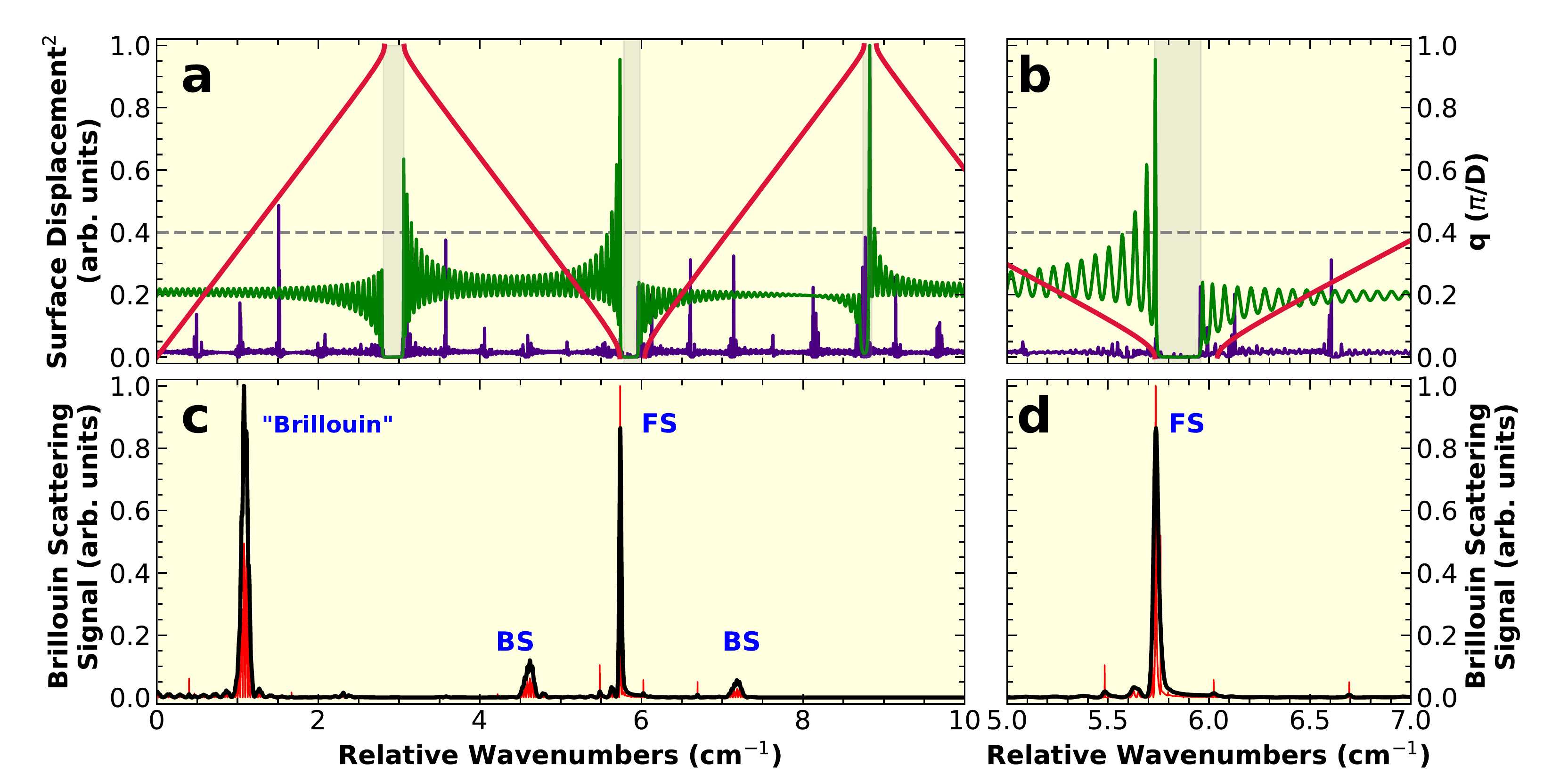}
	\caption{Acoustic surface displacement and Brillouin scattering spectrum. The top panels display the calculated squared acoustic surface displacements for the bare SL (green curve) and for the complete optical microcavity structure with the embedded SL (violet). The tilted infinite SL's dispersion corresponds to the red curve. Bottom panels: The red curve corresponds to the calculated Brillouin spectrum and the black curve indicates its convolution with the experimental resolution.}\label{fig: acoustic displacement and Raman spectra}
\end{center}
\end{figure}

The calculated Brillouin scattering spectrum is shown in Fig.2(d) of the main text and in the bottom panels of Supplementary Fig.\,\ref{fig: acoustic displacement and Raman spectra} (left: extended spectral region, right: detail of the region of interest). The one-dimensional simulation uses classical electrodynamics and a continuum elastic theory and describes all the involved process: incident photon, the creation/annihilation of the acoustic phonons, and the scattered electric field. Details for this well-established method can be found elsewhere \citep{Fainstein-LightScattSolIX(2007), He-PRB37-4086(1988)}. The main ingredient of this calculations is based on the overlap integral of the incoming electric field, the involved acoustic strain field, and the outgoing (scattered) electric field. In Supplementary Fig.\,\ref{fig: strain-EField overlap} we plot the product of these three fields $\eta(z)\mathcal{E}_s^{\ast}(z)\mathcal{E}_i^{\ast}(z)$, for the slow Brillouin zone-center vibrational mode at 5.7\,cm$^{-1}$ ($\sim$180\,GHz). Superimposed, in red, we outline the real part of the refraction index along the direction of the heterostructure, where the different layers can be distinguished. Position ``$0$'' corresponds to the sample's surface. 
The right panel of Supplementary Fig.\,\ref{fig: strain-EField overlap} highlights the detail of the optical cavity spacer with the embedded SL. The fields' product is mainly centred at the cavity and is symmetric with respect to it's centre. It is worth mentioning that the strain field $\eta(z)$ of this acoustic slow Brillouin zone-center mode is the one that maximizes the overlap with the SL's GaAs quantum wells, material that provides the largest photoelastic value, and therefore has the main contribution to the Brillouin spectrum. As observed in Supplementary Fig.\,\ref{fig: strain-EField overlap}, a small contribution will also come from the slight penetration of the photonic and phononic fields into the DBRs. 
\begin{figure}[]
\begin{center}
	\begin{subfigure}{0.49\columnwidth} 
		\includegraphics[width=\textwidth]{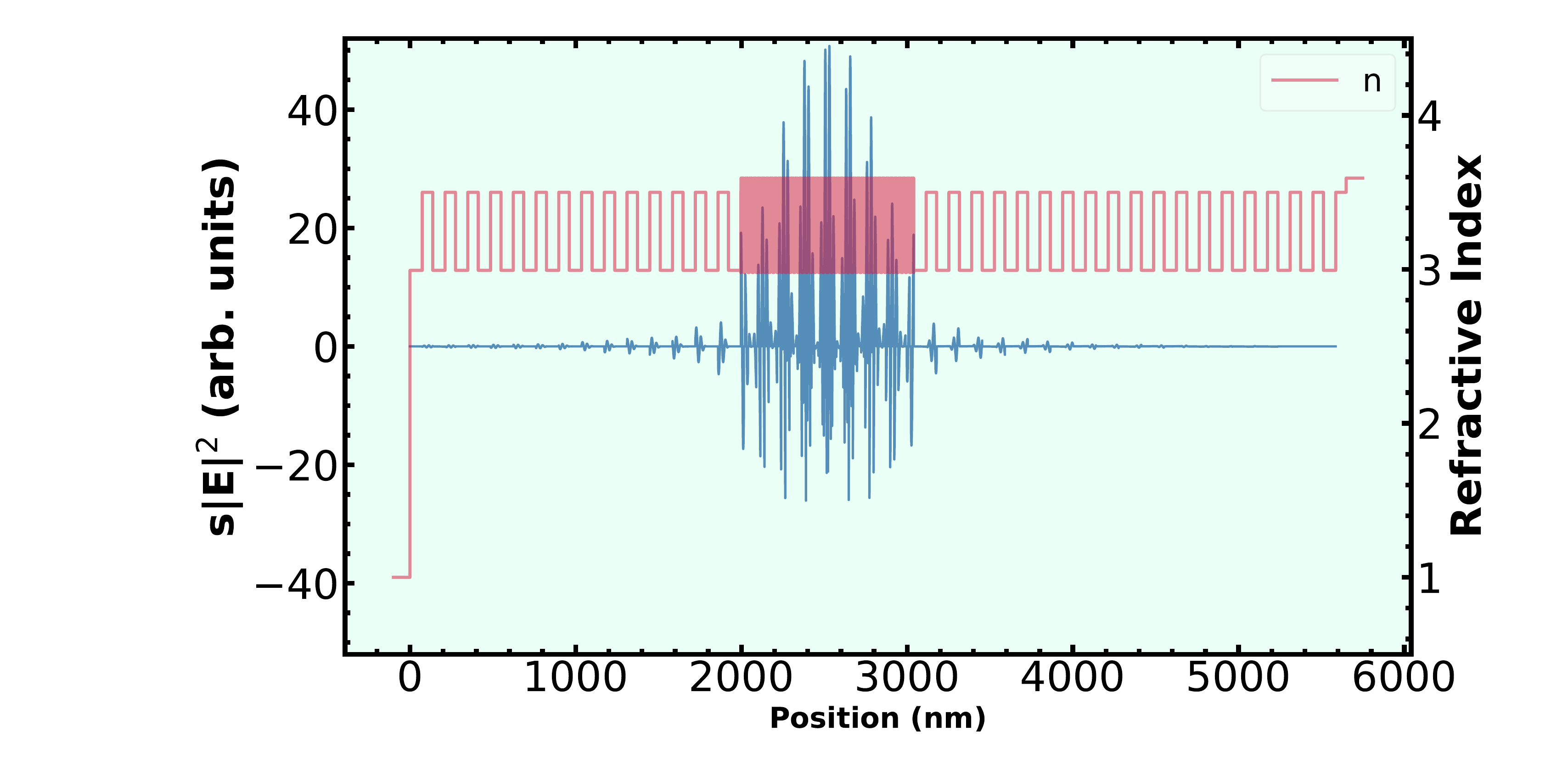}
	\end{subfigure}
	\begin{subfigure}{0.49\columnwidth} 
		\includegraphics[angle=270, width=\textwidth]{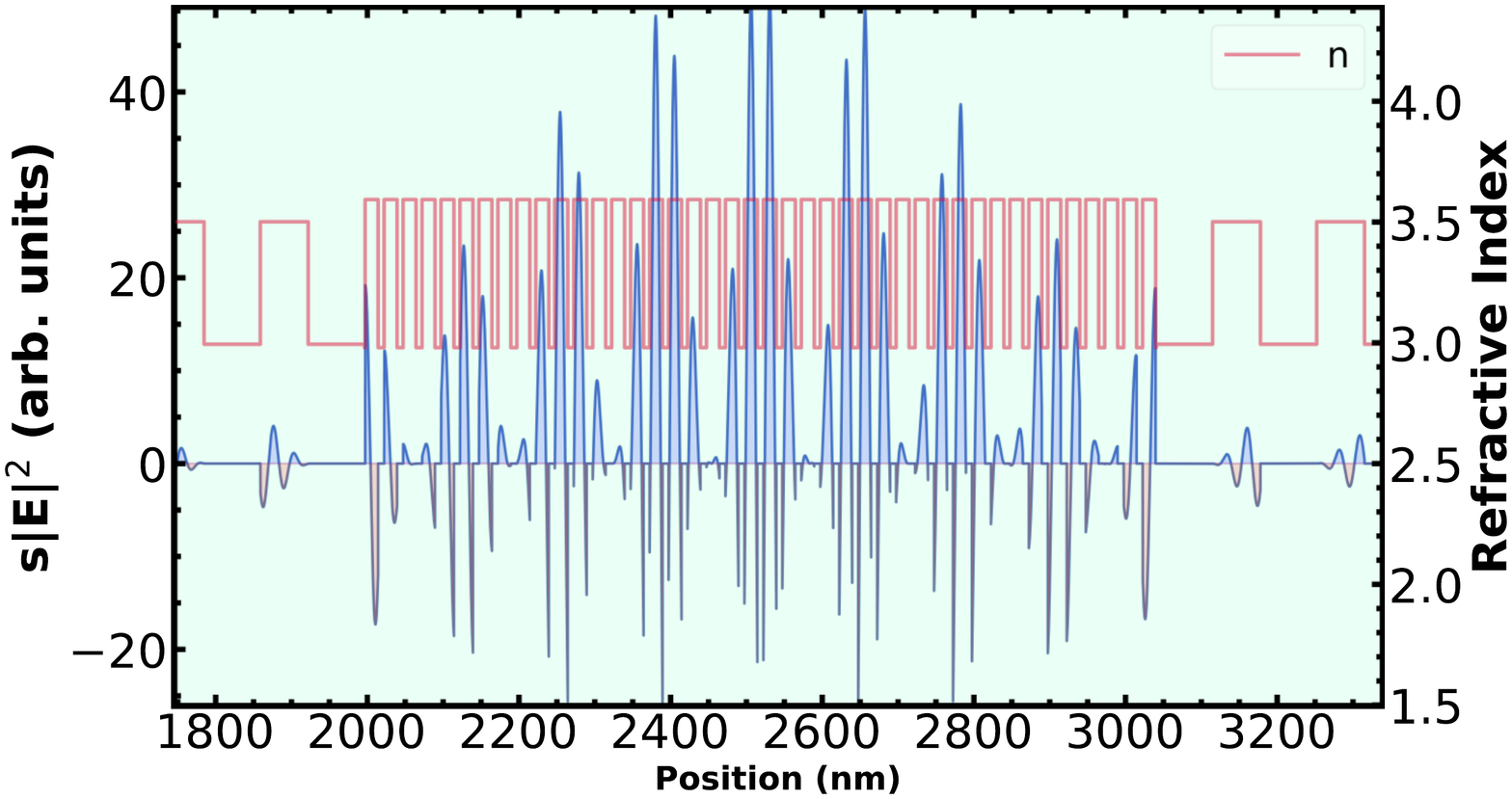}
	\end{subfigure}	
\vspace{-0.7cm}
	\caption{Product of the incoming and scattered electric fields and the acoustic strain field as a function of the heterostructure's position. The acoustic mode corresponds to the slow Brillouin zone-center mode at 5.7\,cm$^{-1}$ (180\,GHz). The superimposed red curves correspond to the real part of the index of refraction along the heterostructure. The right panel shows the detail of the cavity spacer with the embedded SL.}\label{fig: strain-EField overlap}
\end{center}
\end{figure}
The red lines in Supplementary Fig.\,\ref{fig: acoustic displacement and Raman spectra} (bottom panels) are the result of the calculated Brillouin scattering intensity, and the black curve corresponds to this same result but convoluted with the experimental resolution (0.12\,cm$^{-1}$) of the used spectrometer. From this result (see the detail in the right bottom panel) it is quite clear that the measured peaks are resolution limited. 
The overall spectrum reveals the ``Brillouin'' peak at lowest relative energy, together with the well known folded acoustic (FA) triplets near the SL's Brillouin zone-center \cite{Jusserand-LightScattSol5(1989)}. The ``Brillouin'' peak (not observed in our experiments), together with the peaks indicated as BS (back-scattering), correspond to the effective acoustic wavevector $q\simeq 2 k_L$ ($k_L$ is the wavevector of the incident light), and the intense peak at 5.7\,cm$^{-1}$ (indicated as FS) is the slow Brillouin zone-center mode with $q\simeq 0$ (see e.g. Refs.\,[\onlinecite{Jusserand-LightScattSol5(1989)}] and [\onlinecite{Fainstein-LightScattSolIX(2007)}] for details).

\clearpage

\section{Phenomenological model for the Brillouin scattering intensity dependence with TL power}\label{sec: Phenomenological model for the Raman intensity dependence with TL power}

In this section, we will discuss with more detail the phenomenological model used to estimate the evolution of the Brillouin intensity with varying resonant conditions when tuning the trap laser power [red curve in Fig.\,3(b) of the main text]. 
The Brillouin scattering resonances are modelled using peaked functions around the corresponding energies that match the single optical resonances (SOR) or the double optical resonances (DOR). In this case -for simplicity- we have chosen for \textit{each} resonance to use a Gaussian function.\\

The amplitude for each Brillouin scattering process will be proportional to the inverse of the product of the lateral extension (effective width) of the involved incoming ($i$-th) and outgoing ($j$-th) confined optical modes. The Brillouin scattered amplitude $A_{ji}$ of the process, mediated by a phonon from the $i$-th to the $j$-th optical mode ($i\rightarrow j$), will be given by
\begin{eqnarray}\label{eqn: Brillouin scattered amplitude incoming channel i outgoing channel j} 
A_{ji}\propto (D_i\,D_j)^{-1} . 
\end{eqnarray}
Here $D_{\ell}$ represents the effective width of the $\ell$-th mode, and are explained in the \ref{subsec: Laser induced optical potential}. Notice that this amplitude factor [Supplementary Eqn.\eqref{eqn: Brillouin scattered amplitude incoming channel i outgoing channel j}] is of great importance for the calculation of the single-photon coupling rate ($g_0$) [see Eqn.(1) in the main text, and the discussion in the \ref{sec: Single-photon photoelastic coupling rate}]. In fact, the calculation of the Brillouin scattered intensity and the $g_0$ are similar. 
For the case in which $i=j$, the scattering process is considered to be a \textit{single}-mode process. Otherwise, if $i\neq j$ the scattering process is considered to be a \textit{two}-mode process.\\

As pointed out in the main text, the Brillouin scattering process can be resonant with the optical confined modes of the trap. Moreover, the resonance can be considered at single optical resonance or at double optical resonance, and the single resonance can be either incoming or outgoing. For example, if the incident photon is tuned to the fundamental (first) confined mode and the power is such that only this confined mode exists, the amplitude of this Brillouin scattering process will be given by
\begin{eqnarray}
A_{11}\propto (D_1\,D_1)^{-1} . 
\end{eqnarray}
Here the incident photon enters through the optical mode, is scattered, and exits through the residual transmittance of the \textit{same} mode at the energy of one phonon lower. \\

A similar situation is found if the TL power is such that two optical modes are confined within the trap. In this case, the scattering process involves the two modes. That is the incoming photons can enter through one mode, scatter and exit through the same (single-mode process) or the other mode (two-mode process). The example for a process where the incoming channel happens through the 2nd mode and the exit channel goes through the 1st mode, process' amplitude will be given by
\begin{eqnarray}
A_{12}\propto (D_1\,D_2)^{-1} . 
\end{eqnarray}
Following this line of argumentation, we can define the rest of the involved Brillouin scattering amplitudes ($A_{ij}$), depending on the number of modes involved. The effective widths $D_j$ are estimated as described in the \ref{subsec: Laser induced optical potential} and shown in Supplementary Fig.\,\ref{fig: 2a5}.\\
\\

Our system presents five different resonances, which are indicated in Fig.\,3 of the main text and they are tuned when varying the TL power. As mentioned earlier, within this simple phenomenological model, the contribution to the Brillouin scattering intensity is represented by Gaussian functions, whose amplitude is a combination of the involved scattering amplitudes $A_{ji}$. 
In addition, each above described $A_{ji}$ will have a different weight factor depending on the resonant condition, i.e. if the process is in SOR, DOR, or is out of resonance. This weight factor associated with each optical confined mode, will basically be dictated by the coupling probability of the corresponding confined modes to the external continuum of photons. In other words, an external photon will not have the same probability to couple to the photonic modes at exact resonance or slightly detuned from the mode. The coupling probability will be maximal at exact resonance and will decay symmetrically towards either side, basically following the spectral distribution of the corresponding confined optical mode. We will consider that the incoming or outgoing coupling probability of an external photon of energy $\hbar\omega$ to the $j$-th confined mode (of energy $\hbar\omega_j$), is of the form 
\begin{eqnarray}
\mathcal{P}_j(\omega)=\left[1+\left(\frac{\hbar\omega-\hbar\omega_j}{\kappa/2}\right)^2\right]^{-1}\ .
\end{eqnarray}
Here, the same spectral width $\kappa$ is considered for each mode.\\

For example, for the case of the first SOR situation in Fig.\,3 of the main text ($P_{\textrm{TL}}\sim$6.25\,mW), only one confined mode exists with energy $\hbar\omega_1=\hbar\omega_{\textrm{BL}}$ (see Fig.3a of the main text). We are in the presence of a single-mode incoming SOR. This is interpreted as the probability $\mathcal{P}_1(\omega_{1})$ of entering the trap through the first (fundamental) mode, and the probability $\mathcal{P}_1(\omega_{1}-\Omega)$ of exiting the trap out-of-resonance after the Stokes process, with an energy that is $\hbar\Omega$ (one phonon quanta) less. The joint probability of the process as a whole will thus be the product of both probabilities. The intensity of the corresponding Brillouin process, that corresponds to the amplitude of the associated Gaussian function  will hence be given by 
\begin{eqnarray}
G_1=\underbrace{\mathcal{P}_1(\omega_{1})}_{1}\,\underbrace{\mathcal{P}_1(\omega_{1}-\Omega)}_{\sim 1/24}~A_{11}\simeq \mbox{$\frac{1}{24}$}~ A_{11}\ .
\end{eqnarray}

For the case of the second SOR, indicated in Fig.\,3 of the main text ($P_{\textrm{TL}}\sim$8.75\,mW and where $\omega_2=\omega_{\textrm{BL}}$), two optical modes are involved. Consequently, the following processes are possible: (\textit{i}) the resonant incoming photon is coupled resonantly to the mode 2 with $\mathcal{P}_2(\omega_{2})$, (\textit{ii}) is scattered by one phonon, and  (\textit{iii}) exits the optical non-resonantly either through the same mode 2 or the mode 1. This will have associated probabilities $\mathcal{P}_2(\omega_{2}-\Omega)$ or $\mathcal{P}_1(\omega_{2}-\Omega)$, respectively. 
Less probable is the process in which the photon enters through the mode 1 non-resonantly with $\omega_2$, and exits non-resonantly through the same out-coming channels as before. These double non-resonant processes, expected to be small, are neglected in what follows.\\
Since the joint probabilities of each incoming-outcoming channel pairs are mutually exclusive events (if one happens, the other does not), the probabilities with their respective Brillouin amplitude weights are simply the sum. The contribution to the Brillouin intensity of the second SOR will thus be
\begin{eqnarray}
 G_{2}&=&\overbrace{\underbrace{\mathcal{P}_2(\omega_{2})}_{1}\,\underbrace{\mathcal{P}_1(\omega_{2}-\Omega)}_{\sim 1/9}~A_{12}+\underbrace{\mathcal{P}_2(\omega_{2})}_{1}\,\underbrace{\mathcal{P}_2(\omega_{2}-\Omega)}_{\sim 1/24}~A_{22}}^{in-SOR}+\cancel{\overbrace{\underbrace{\mathcal{P}_1(\omega_{2})}_{\sim 1/5}\underbrace{\mathcal{P}_1(\omega_{2}-\Omega)}_{\sim 1/9}~A_{11}+\underbrace{\mathcal{P}_1(\omega_{2})}_{\sim 1/5}\underbrace{\mathcal{P}_2(\omega_{2}-\Omega)}_{\sim 1/24}~  A_{12}}^{non-res}}\nonumber\\
&=&\mbox{$\frac{1}{9}$}A_{12}+\mbox{$\frac{1}{24}$}~ A_{22}
\end{eqnarray}

Following the above procedure, we can extend these arguments for the case in which more optical confined modes are involved. In particular, for the three DORs indicated in Fig.\,3(a) of the main text, namely the one resonantly entering through the 3rd confined mode and resonantly outgoing through the 1st confined mode ($3\rightarrow 1$), the one entering the 4th mode and resonantly outgoing through the 2nd mode ($4\rightarrow 2$), and last, the one entering resonantly through the 5th optical confined mode and outgoing at resonance with the 3rd mode ($5\rightarrow 3$), the corresponding Gaussian intensities for each process will have the following form, respectively 
\begin{eqnarray}
&&G_{\text{\tiny 3$\rightarrow$1}}==A_{13}+ \mbox{$\frac{1}{5}$}~A_{12}+\mbox{$\frac{1}{7}$} ~A_{14}+\mbox{$\frac{1}{7}$}~A_{43}+\mbox{$\frac{1}{8}$}~A_{23}+\mbox{$\frac{1}{24}$}~A_{11}+\mbox{$\frac{1}{24}$}~A_{33}+\mbox{$\frac{1}{35}$}~A_{42}+\mbox{$\frac{1}{40}$}~A_{22}+\mbox{$\frac{1}{49}$}~A_{44}\\
&&G_{\text{\tiny 4$\rightarrow$2}}=A_{24}+\mbox{$\frac{1}{6}$}~A_{34}+\mbox{$\frac{1}{8}$}~A_{32}+ \mbox{$\frac{1}{16}$}~A_{14}+\mbox{$\frac{1}{24}$}~A_{44}+\mbox{$\frac{1}{27}$}~A_{22}+ \mbox{$\frac{1}{48}$}~A_{33}\\
&&G_{\text{\tiny 5$\rightarrow$3}}=A_{35}+\mbox{$\frac{1}{6}$}~A_{34}+\mbox{$\frac{1}{7}$}~A_{45}+\mbox{$\frac{1}{11}$}~A_{25}+\mbox{$\frac{1}{24}$}~A_{55}+\mbox{$\frac{1}{27}$}~A_{33}+\mbox{$\frac{1}{49}$}~A_{44}\ ,
\end{eqnarray}
where we also have neglected the non-resonant weakly contributing terms. \\

If a non-coherent sum of the resonant Brillouin scattering processes is considered, we add all five Gaussian functions with the above amplitudes. Normalizing against $G_{\text{\tiny 3$\rightarrow$1}}$, which is the process that has the largest intensity, we obtain the red continuous curve displayed in Fig.\,3(b) of the main text. We used a standard Gaussian deviation of 1.5\,meV for all optical resonances.

\clearpage

\section{photoelastic parameters}\label{sec: Considered photoelastic parameters}

The pertinent photoelastic element for the experiments described in this work corresponds to $p_{12}$ (using the compact tensor notation). At room temperature and for $\Delta = -37.7\,$meV (see in the \ref{sec: Sample details}, Supplementary Fig.\,\ref{FigS02}),  $p_{12}$ reaches a value of $\sim$0.61 for GaAs\,\cite{Jusserand}. 
For Al$_{0.1}$Ga$_{0.9}$As, since the Gallium content is high, we can estimate the value for the photoelastic constant from the data in Ref.\,[\onlinecite{Jusserand}], considering that its band-gap has a shift of $\sim$124\,meV due to the Aluminium content of the alloy. Thus, for the latter $p_{12}\simeq$\,0.20. This value has important implications, since the photoelastic optomechanical interaction of the photonic and phononic fields must be taken into account in the whole structure, and not only within the spacer. The photoelastic contribution for AlAs and Al$_{0.95}$Ga$_{0.05}$As, are negligible for the experimental situation considered here and are assumed to be zero.

\section{Single-photon photoelastic coupling rate ($g_0$)}\label{sec: Single-photon photoelastic coupling rate}

The main contribution to the photon-acoustic phonon coupling for near excitonic resonant condition is photoelastic \cite{Villafane}. The single-photon photoelastic coupling rate of an acoustic phonon mode with a given frequency $\Omega_m$ is estimated from the overlap integral that involves the corresponding stain field $\eta_m(z)$, and the incident and scattered electric fields, $E_{\textrm{i}}$ and $E_{\textrm{s}}$, respectively. Interaction Hamiltonian for the photoelastic contribution to the optomechanical coupling is given by \cite{Kharel-ScienceAdvances5-eaav0582(2019), Rakich-OpticsExpress18-14439(2010)}
\begin{eqnarray}\label{eqn: single-mode coupling Hamiltonian}
\mathcal{H}^{\textrm{int}}= \mbox{$\frac{1}{2}$}\int_V dV \,\epsilon_\textrm{o}\, \epsilon_\textrm{r}^2\,p^{}_{ijkl}\, \eta^{}_{kl}\,E_i^{\ast}\,E_j^{}\ .
\end{eqnarray}
$\epsilon_\textrm{o}$ is the vacuum dielectric permitivity, $\epsilon_{\textrm{r}}$ is the relative dielectric function, $p_{ijkl}$ represents the photoelastic tensor, $\eta^{}_{kl}$ the strain tensor, and $E$  the electric field.\\

For the case of only \textit{two} optical confined optical modes, as is the case of our experimental situation, in terms of the normal electric fields and acoustic displacement modes, the above interaction Hamiltonian that couples to one phonon mode yields \cite{Kharel-ScienceAdvances5-eaav0582(2019)}
\begin{eqnarray*}
\mathcal{H}^{\textrm{int}}=-\hbar\,g_0\,\big[\hat{a}^{\dagger}_{\textrm{s}}\,\hat{a}^{}_{\textrm{i}}\,\hat{b}^{}+\textrm{H.c.}\big]\ .
\end{eqnarray*}
$\hat{a}^{\dagger}_{j}$($\hat{a}^{}_j$) and $\hat{b}^{\dagger}$($\hat{b}^{}$) are the photon and acoustic-phonon modes' creation(destruction) operators, respectively. The indexes $\textrm{i}$($\textrm{s}$) stand for the incident(scattered) photons. The single-photon photoelastic coupling rate is given by \cite{Villafane, Kharel-ScienceAdvances5-eaav0582(2019)}
\begin{eqnarray}\label{eqn: single-photon photoelastic coupling 1}
g_0&=&\mbox{$\frac{1}{\hbar}$}\,\int_V \epsilon_o\,\epsilon_{\textrm{r}}^2(z)\,p_{12}(z)\,\eta(z)\,E^{\ast}_{\omega_{\textrm{s}}}(z)\,E^{}_{\omega_{\textrm{i}}}(z)\,dV\ .
\end{eqnarray}

\subsection*{Phonon mode and associated strain field}\label{subsec: Phonon mode and associated strain field}

Here we assume a one-dimensional continuum elastic problem, supported by the fact that the changes of the involved acoustic parameters, due to the trap-laser induced temperature modification, are small. The longitudinal acoustic displacement field ($\textrm{u}_z$) in the $z$-direction is obtained by solving the elastic acoustic wave equation for the entire heterostructure\,\cite{Landau_Book-Elasticity(1986)}
\begin{eqnarray*}
\rho\,\ddot{\textrm{u}}_z=\partial_z [C_{33}\, \eta_{zz}]\ ,
\end{eqnarray*}
where $\eta_{zz}$ is the corresponding strain tensor component, and $C_{33}$ is the pertinent stiffness tensors element. \\

The formal solution is found using the transfer-matrix formalism \cite{Fainstein-LightScattSolIX(2007)}, imposing stress-free boundary conditions (b.c.) at both sample's ends (surface and back-side of the substrate). The following expression for the spatial eigenmode of frequency $\Omega_m$ is obtained
\begin{eqnarray*}
u_m(z)=a_m\,e^{i\,q_m\,z}+b_m\,e^{-i\,q_m\,z}\ ,
\end{eqnarray*}
where $q_m=\Omega_m/v_{ac}$ is the corresponding wavevector within each layer, and the coefficients $a_m$ and $b_m$ are defined by the b.c. and by the Sturm-Liouville normalization condition
\begin{eqnarray*}
&&\int_V u^{\ast}_m\, \rho(z)\,u^{}_m\,dV=1 \quad\Rightarrow\\
&&\Rightarrow\quad \int_{L} \tilde{u}^{\ast}_m(z)\, \rho(z)\,\tilde{u}^{}_m(z)\,dz=1 \ .
\end{eqnarray*}
Here we have defined $u^{}_m(z)=\tilde{u}^{}_m(z)/\sqrt{A_{\textrm{ac}}}$, and where $A_{\textrm{ac}}$ corresponds to the effective transverse area of the acoustic mode. $L$ indicates the integration interval along the entire heterostructure.\\

The general acoustic displacement $\textrm{u}(z,t)$ will be given by \cite{Kharel-ScienceAdvances5-eaav0582(2019)}
\begin{eqnarray*}
\textrm{u}(z,t)=\sum_m r_m\,\frac{\tilde{u}_m(z)}{\sqrt{A_{\textrm{ac}}}}\,\big[\hat{b}^{}_m(t)+\hat{b}^{\dagger}_m(t)\big]\ ,
\end{eqnarray*}
where the amplitude comes from the quantification and results $r_m=\sqrt{\frac{\hbar}{2\,\Omega_m}}$.\\

The corresponding strain field for the $m$-th mode is straightforward, namely 
\begin{eqnarray}\label{eqn: strain field}
\eta_m(z,t)&=&\partial_z\textrm{u}_m(z,t)\nonumber\\
&=&\mbox{$\sqrt{\frac{\hbar}{2\,A_{\textrm{ac}}\,\Omega_m}}$}\,\partial_z\tilde{u}_m(z)\,\big[\hat{b}^{}_m(t)+\hat{b}^{\dagger}_m(t)\big]\ .
\end{eqnarray}

The above phonon operators are given in the Heisenberg representation.

\subsection*{Electric fields}\label{subsec: Electric fields}

The involved electric fields, correspond to the optical cavity modes that are well confined within the optical spacer, and decay \textit{exponentially} towards the top and bottom DBRs. The normalization of the electric field has been a long standing issue. We will consider the same approach as the one suggested by Prashanta Kharel, \textit{et al.} \cite{Kharel-ScienceAdvances5-eaav0582(2019)}. Since the fields are mainly localized within the spacer and part of the DBRs, the normalization is estimated by considering the structure as an effective dielectric medium $\epsilon_{\textrm{r}}^{\text{\tiny eff}}$ and weighted by the field's effective volume $V_{\textrm{opt}}^{\textrm{\tiny eff}}= A^{}_{\textrm{opt}}\,L_{\textrm{opt}}^{\text{\tiny eff}}$. $L_{\textrm{opt}}^{\text{\tiny eff}}$ corresponds to the penetration of the mode within the DBRs \cite{Babic-IEEE-JoQE28-514(1992)} added to the $\frac{9}{2}\lambda$-spacer thicknes. For the system considered in this work we estimate $L_{\textrm{opt}}^{\text{eff}}\sim5\mu$m. 

Since the interaction with the 1D phonons acts only in the stacking direction of the structure, and given the experimental incidence and collection, which is normal to the sample's surface, added to the relative high dielectric contrast between vacuum and the dielectric media, we approximate the in-plane polarized fields assuming a one-dimensional problem. The mode's spatial profile $\mathcal{E}_\omega(z)$ is obtained by solving Maxwell's equations and the linearly polarized electromagnetic 1D field's wave equations, with the corresponding boundary conditions for non-magnetic materials, and using a transfer-matrix approach setting the incident amplitude to unity. Thus, the obtained fields are of the form \cite{Kharel-ScienceAdvances5-eaav0582(2019),Villafane}
\begin{eqnarray}
&&E_{\omega}(z)=\sqrt{\mbox{$\frac{\hbar\omega}{2\,\epsilon_o\epsilon_{\textrm{r}}^{\text{eff}}\,A_{\textrm{opt}}\,L_{\textrm{opt}}^{\text{eff}}}$}}\,\mathcal{E}_\omega(z)\,\big[\hat{a}^{}_\omega(t)+\hat{a}^{\dagger}_\omega(t)\big]\ ,\label{eqn: el field}\\
&&\quad\mathcal{E}_\omega(z)=A_{\omega}\,e^{i k_\omega z}+B_{\omega}\,e^{-i k_\omega z}\ .\nonumber
\end{eqnarray}
Within each layer, the coefficients $A_{\omega}$ and $B_{\omega}$ result from propagating the corresponding transfer matrix, and $k_\omega$ is the wavevector associated to the  photon of energy $\hbar\omega$. \\

It can be shown that the field distribution inside the cavity along the heterostructure for the different modes is very similar. In addition, considering that the phonon frequency is much smaller than that of the involved electric fields, we will adopt the same profile for the variation along the heterostructure's direction for both the incident and scattered field modes.\cite{Fainstein-LightScattSolIX(2007)}

\subsection*{Photoelastic coupling rate}
Replacing Supplementary Eqns.\eqref{eqn: strain field} and \eqref{eqn: el field} in expression \eqref{eqn: single-photon photoelastic coupling 1}, integrating over the in-plane direction (with an effective area $\bar{A}$), we obtain the \textit{single-photon photoelastic coupling rate} with the phononic mode \cite{Kharel-ScienceAdvances5-eaav0582(2019), Villafane}
\begin{eqnarray}\label{eqn: g_0}
g_0=\mbox{$\sqrt{\frac{\hbar}{2\,A_{\textrm{ac}}\,\Omega_{\textrm{m}}}}$}\,\mbox{$\frac{\sqrt{\omega_{\textrm{s}}\,\omega_{\textrm{i}}}}{2\,\epsilon_{\textrm{r}}^{\text{eff}}\,L_{\textrm{opt}}^{\text{eff}}}$} \mbox{$\frac{\bar{A}}{\sqrt{A_{\textrm{opt}}^{\textrm{i}}\,A_{\textrm{opt}}^{\textrm{s}}}}$}\!\int_{L} \!\epsilon_{\textrm{r}}^2(z)\,p_{12}(z)\partial_z\tilde{u}_{\textrm{m}}(z)\,\mathcal{E}^{\ast}_{\omega_{\textrm{s}}}(z)\,\mathcal{E}^{}_{\omega_{\textrm{i}}}(z)\,dz\ .
\end{eqnarray}

The effective dielectric function is obtained as $\epsilon_{\textrm{r}}^{\text{eff}}=d_{\textrm{T}}^{-1}\sum_j \epsilon_j\,d_j$, and $d^{}_{\textrm{T}}$ is the total structure's thickness and $d_j$ the width of the $j$-th layer of the heterostructure.\\

The value for $A_{\textrm{ac}}$ is assumed to be equal to $\bar{A}$. This value will depend mainly on the lateral distribution of the two electric field modes, and will be therefore accounted for as the product of both transverse characteristic transverse dimensions, or equivalently the product of the corresponding areas, that is $\bar{A}\simeq (A_{\textrm{opt}}^{\textrm{i}}\,A_{\textrm{opt}}^{\textrm{s}})^{-1}$. Consequently, Supplementary Eqn.\eqref{eqn: g_0} reduces to
\begin{eqnarray}
g_0=\mbox{$\frac{1}{2\,\epsilon_{\textrm{r}}^{\text{eff}}\,L_{\textrm{opt}}^{\text{eff}}}$}\mbox{$\sqrt{\frac{\hbar\,\omega_{\textrm{s}}\,\omega_{\textrm{i}}}{2\,\Omega_{\textrm{m}}}}$} \underbrace{\mbox{$\frac{1}{(A_{\textrm{opt}}^{\textrm{i}}\,A_{\textrm{opt}}^{\textrm{s}})^{1/4}}$}}_{\propto1/\sqrt{D_{\textrm{i}}\,D_{\textrm{s}}}}\!\int_{L} \!\epsilon_{\textrm{r}}^2(z)\,p_{12}(z)\partial_z\tilde{u}_{\textrm{m}}(z)\,\mathcal{E}^{\ast}_{\omega_{\textrm{s}}}(z)\,\mathcal{E}^{}_{\omega_{\textrm{i}}}(z)\,dz\ .
\end{eqnarray}
$D_{\textrm{i}}$($D_{\textrm{s}}$) is the characteristic diameter of the incident(scattered) confined electric mode. \\

We finally obtain Eqn.(1) of the main text
\begin{eqnarray}
g_0=\mathcal{K} \mbox{$\frac{1}{\sqrt{D_{\textrm{i}}\,D_{\textrm{s}}}}$}\!\int_{L} \!\epsilon_{\textrm{r}}^2(z)\,p_{12}(z)\partial_z\tilde{u}_{\textrm{m}}(z)\,\mathcal{E}^{\ast}_{\omega_{\textrm{s}}}(z)\,\mathcal{E}^{}_{\omega_{\textrm{i}}}(z)\,dz \ ,
\end{eqnarray}
where $\mathcal{K}=\frac{1}{2\,\epsilon_{\textrm{r}}^{\text{eff}}\,L_{\textrm{opt}}^{\text{eff}}}\sqrt{\frac{4\,\hbar\,\omega_{\textrm{s}}\,\omega_{\textrm{i}}}{2\,\pi\,\Omega_{\textrm{m}}}}$.\\

Using the usual acoustic and dielectric room temperature parameters for this AlGaAs system, and for the photoelastic parameters described in \ref{sec: Considered photoelastic parameters}, we obtain the reported single-photon photoelastic coupling rate for the $\sim$180\,GHz superlattice mode of  
\begin{eqnarray*}
g_0= 2\pi\times 1.7\,\textrm{MHz}\ .
\end{eqnarray*}


\end{document}